 \documentclass[iop]{emulateapj}
 \usepackage{color}
 \usepackage{natbib}

 \def \ms{m\,s$^{-1}$\,}

\begin{document}
 \shorttitle{Neptune-mass planet orbiting GJ 687}
 \shortauthors{Burt et al.}

 \title{The Lick-Carnegie Exoplanet Survey:
 Gliese 687~\MakeLowercase{b}: A Neptune-Mass Planet Orbiting a Nearby Red Dwarf}
 \author{Jennifer Burt \altaffilmark{1},
 Steven S. Vogt \altaffilmark{1},
 R. Paul Butler \altaffilmark{2},
 Russell Hanson\altaffilmark{1},
 Stefano Meschiari \altaffilmark{3},
 Eugenio J. Rivera \altaffilmark{1},
 Gregory W. Henry \altaffilmark{4},
 Gregory Laughlin \altaffilmark{1}
 }

 \altaffiltext{1}{UCO/Lick Observatory, Department
 of Astronomy and Astrophysics,
 University of California at Santa Cruz, Santa Cruz, CA 95064}
 \altaffiltext{2}{Department of Terrestrial Magnetism,
 Carnegie Institute of Washington, Washington, DC 20015}
 \altaffiltext{3}{McDonald Observatory, University of Texas at Austin, Austin, TX 78752}
 \altaffiltext{4}{Center of Excellence in Information Systems,
 Tennessee State University, Nashville, TN 37209}

 \begin{abstract}
 Precision radial velocities from the Automated Planet Finder and Keck/HIRES reveal an $M\sin(i)=18\pm2\,M_{\oplus}$
 planet orbiting the nearby M3V star GJ~687. This planet has an orbital period,
 $P=38.14$ days, and a low orbital eccentricity. Our Str{\"{o}}mgren b and y photometry of the host star suggests a stellar rotation signature with a period of $P$ = 60 days. The star is somewhat chromospherically active, with a spot filling factor estimated to be several percent. The rotationally--induced 60-day signal, however, is well-separated from the period of the radial velocity variations, instilling confidence in the interpretation of a Keplerian origin for the observed velocity variations. Although GJ~687~b produces relatively little specific interest in connection with its individual properties, a compelling case can be argued that it is worthy of remark as an {\it eminently typical}, yet at a distance of 4.52~pc, a very nearby representative of the galactic planetary census. The detection of GJ~687~b indicates that the APF telescope is well suited to the discovery of low-mass planets orbiting low-mass stars in the as-yet relatively un-surveyed region of the sky near the north celestial pole.

 \end{abstract}


 \section{Introduction}

 The Copernican principle implies that the Earth, and, by extension, the solar system, do not hold a central or specifically favored position. This viewpoint is related to the so-called {\it mediocrity principle} \citep{Kukla10}, which notes that an item drawn at random is more likely to come from a heavily populated category than one which is sparsely populated.

 These principles, however, have not had particularly apparent success when applied in the context of extrasolar planets. \citet{Mayor09} used their high precision Doppler survey data to deduce that of order 50\% (or more) of the chromospherically quiet main-sequence dwarf stars in the solar neighborhood are accompanied by a planet (and in many cases, by multiple planets) with $M\sin(i)\lesssim30M_{\oplus}$, and orbital periods of $P<100\,{\rm d}$. Taken strictly at face value, this result implies that our own solar system, which contains nothing interior to Mercury's $P=88\,{\rm d}$ orbit, did not participate in the galaxy's dominant mode of planet formation. Yet the eight planets of the solar system have provided, and continue to provide, the {\it de-facto} template for most discussions of planet formation.

\begin{figure}
 \plotone{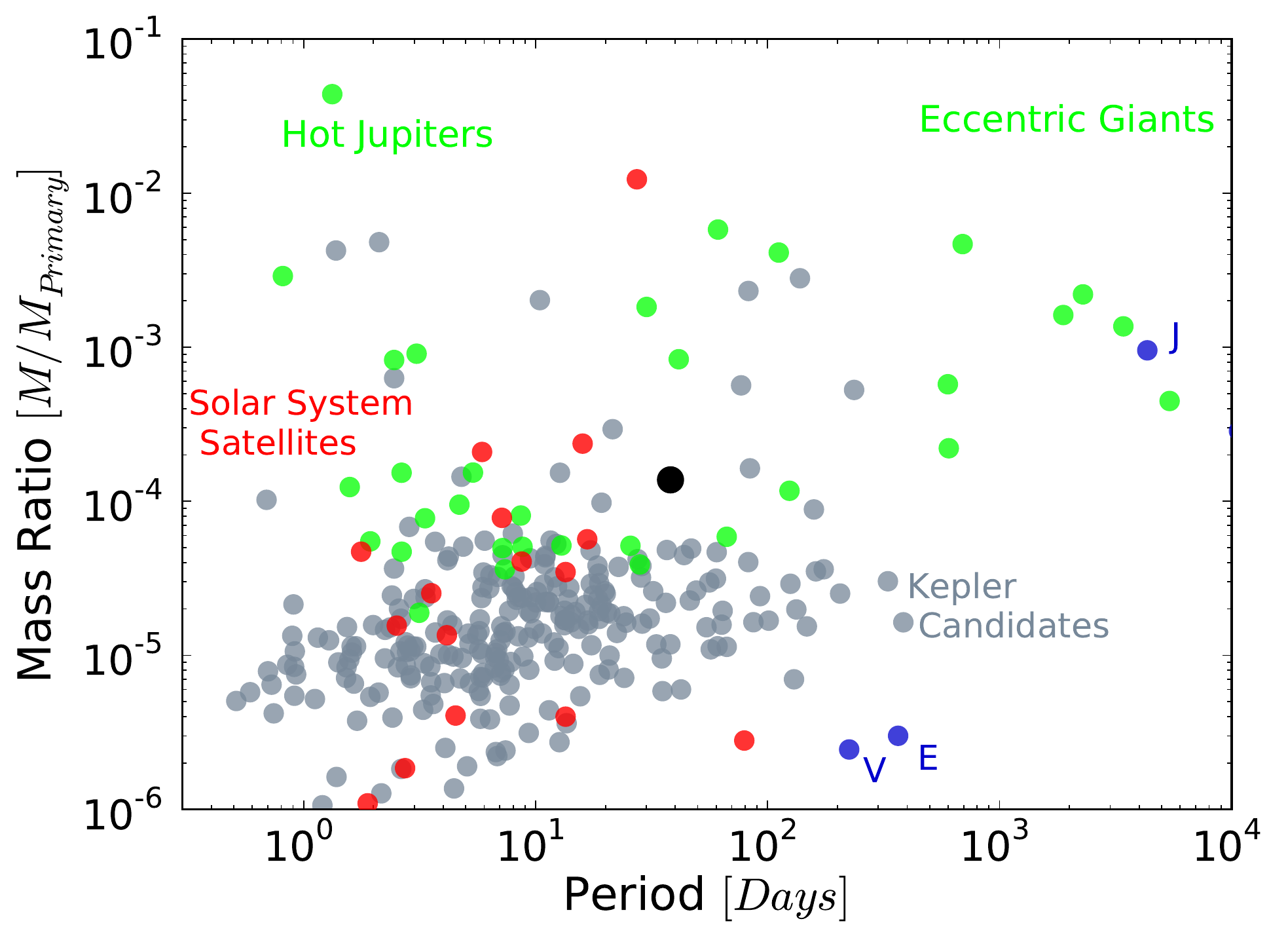}
 \caption{Population digram for currently known extra solar planets orbiting stars with reported masses $M_{{\rm star}} < 0.6 M_{\odot}$. {\it Green circles:} Planets securely detected by the radial velocity method (either with or without photometric transits). {\it Red circles:} The regular satellites of the Jovian planets in the Solar System. {\it Gray circles:} Kepler candidates and objects of interest. Radii for these candidate planets, as reported in \citep{Batalha13}, have been converted to masses assuming $M/M_{\oplus}=(R/R_{\oplus})^{2.06}$ \citep{Lissauer11}, which is obtained by fitting the masses and radii of the solar system planets bounded in mass by Venus and Saturn, which may be a rather naive transformation given the startling range of observed radii for planets with masses between Earth and Uranus. Venus, Earth, and Jupiter are indicated on the diagram for comparison purposes. Data are from www.exoplanets.org, accessed 01/12/2014.}
 \label{fig:mdwarfCensus}
 \end{figure}

 Indeed, where extrasolar planets are concerned, M-dwarfs and mediocrity appear to be effectively synonymous. Recent observational results suggest that low-mass planets orbiting low-mass primaries are by no means rare. Numerous examples of planets with $M_{\rm p}<30\,M_{\oplus}$ and M-dwarf primaries have been reported by the Doppler surveys (e.g. \citealt{Butler04, Mayor09}, and many others), and the {\it Kepler} Mission has indicated that small planets are frequent companions to low mass stars. For example, \citet{Dressing13} report that among dwarf stars with $T_{\rm eff}<4000\,{\rm K}$, the occurrence rate of $0.5\,R_{\oplus}<R_{\rm p}<4\,R_{\oplus}$ planets with $P<50\,{\rm d}$ is $N=0.9^{+0.04}_{-0.03}$ planets per star. Improved statistics, however, are required for a definitive statement that is couched in planetary masses as well as in planetary radii. Figure \ref{fig:mdwarfCensus} shows the current distribution of reported planets and planetary candidates orbiting primaries with $M_{\star}<0.6M_{\odot}$, which we adopt as the functional border between ``M-type'' stars and ``K-type stars''.

 The census of low-mass planets orbiting low-mass primaries can be accessed using a variety of techniques. For objects near the bottom of the main sequence, it appears that transit photometry from either ground \citep{Charbonneau10} or space \citep{Triaud13} offer the best prospects for planetary discovery and characterization. For early to mid M-type dwarfs, there is a large enough population of sufficiently bright primaries that precise Doppler detection (see, e.g. \citealp{Rivera10}) can play a lead role. For the past decade, we have had a sample of $\sim$160 nearby, photometrically quiet M-type stars under precision radial velocity surveillance with the Keck telescope and its
 HIRES spectrometer. In recent months, this survey has been supplemented by data from the Automated Planet Finder Telescope \citep{Vogt14a}. Here, we present 16.6 years of Doppler velocity measurements for the nearby M3 dwarf GJ~687 (including 122 velocity measurements from Keck, 20 velocity measurements from the APF, and 5 velocity measurements made with the Hobby-Eberly Telescope) and we report the detection of the exoplanet that they imply. We use this discovery of what is a highly archetypal representative of a planet in the Milky Way -- in terms of its parent star, its planetary mass, and its orbital period -- to motivate a larger discussion of the frequency of occurrence, physical properties, and detectability of low-mass planets orbiting M-type stars.

 The plan for this paper is as follows. In \S 2, we describe the physical and spectroscopic properties of the red dwarf host star Gliese 687. In \S 3, we describe our radial velocity observations of this star. In \S 4 we describe our Keplerian model for these observations, along with an analysis that assesses our confidence in the detection. In \S 5, we describe our photometric time series data for the star, which aids in the validation of the planet by ruling out spot-modulated interpretations of the Doppler variations. In \S 6, we discuss the ongoing refinement of the planet-metallicity correlation for low-mass primaries, in \S 7, we discuss the overall statistics that have emerged from more than 15 years of precision Doppler observations of M-dwarf stars with the Keck Telescope, and in \S 8 we conclude with an overview that evaluates the important future role of the APF telescope in precision velocimetry of nearby, low-mass stars.

 \section{GJ~687 Stellar Parameters}

\begin{figure}
 \plotone{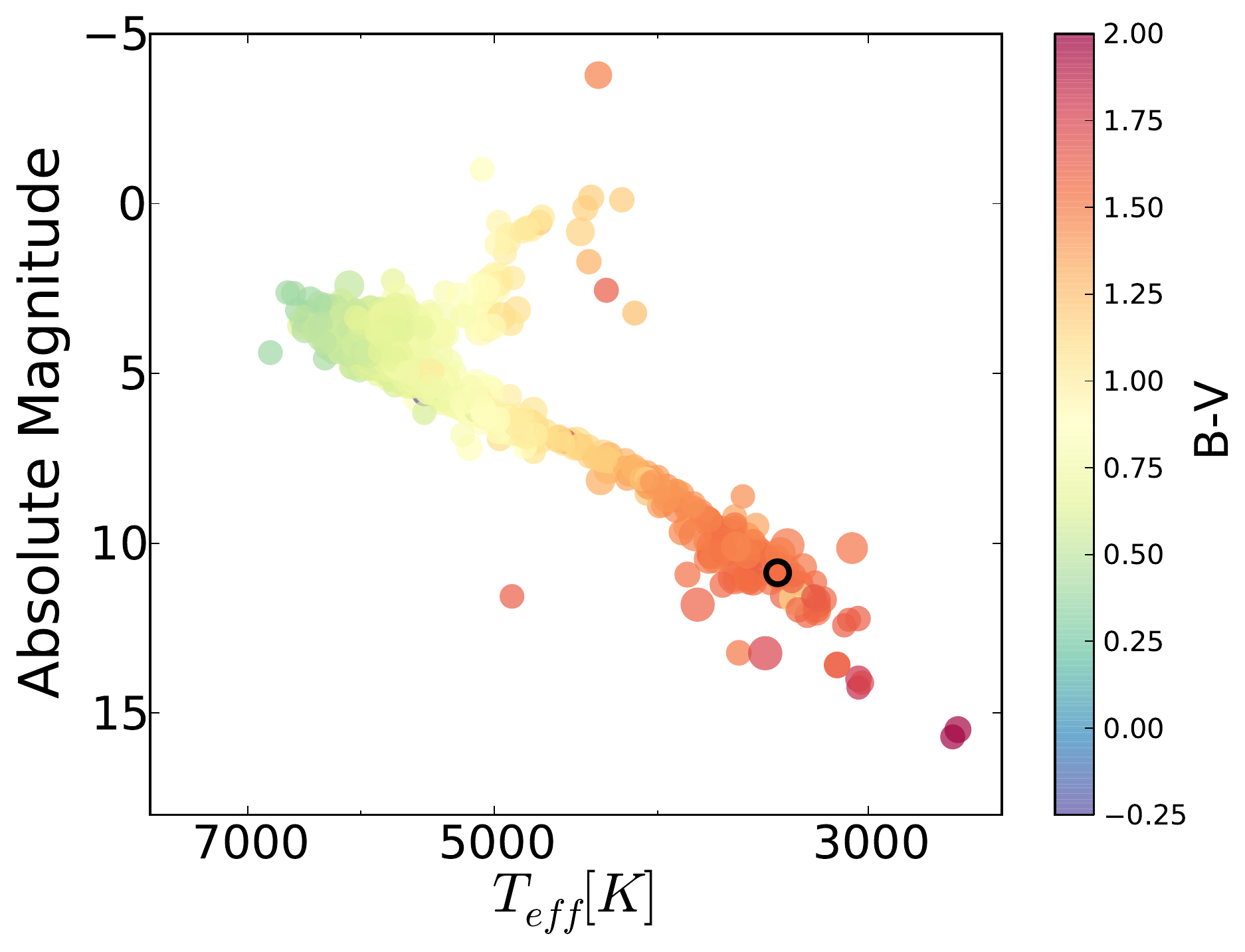}
 \caption{HR diagram with GJ~687's position indicated as a small open circle. Absolute magnitudes, $M$, are estimated from V band apparent magnitudes and Hipparcos distances using $M=V+5\log_{10}(d/10~{\rm pc})$. All 956 stars in our catalog of radial velocity measurements (for which more than 20 Doppler measurements exist) are shown, color-coded by their B-V values, with point areas sized according to the number of observations taken.}
 \label{fig:HRDiag}
 \end{figure}

\begin{figure}
 \plotone{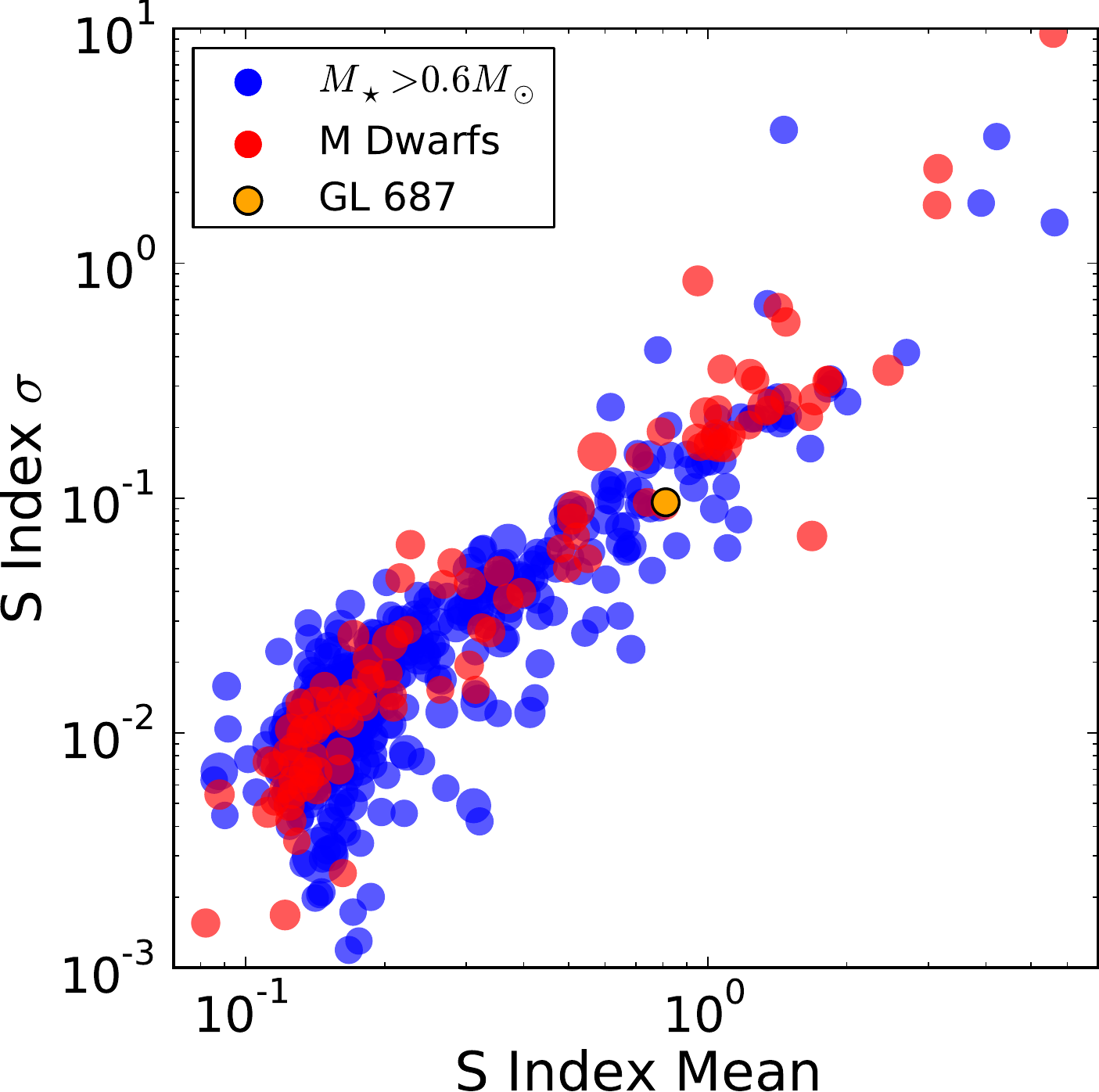}
 \caption{The average value of the S-index against the standard deviation of the S-index for all the stars in the Lick-Carnegie database. Stars with $M_{\star} < 0.6\, M_{\odot}$ are colored red. GJ~687 is shown as an orange circle in the midst of this population, showing that it is a somewhat active star. The areas subtended by the individual points are, in all cases, proportional to the number of Doppler velocity observations that we have collected of the star (with systems above an upper bound of 250 observations receiving the same point size).}
 \label{fig:sindex}
 \end{figure}

 Gliese 687 (LHS 450, BD+68$^{\circ}$946) lies at a distance, $d=4.5$ pc, is the 39th-nearest known stellar system, and is the closest star north of $+60^{\circ}$ declination. Figure \ref{fig:HRDiag} indicates GJ~687's position in the color-magnitude diagram for stars in the Lick-Carnegie Survey's database of Keck observations. Due to its proximity and its brightness (V=9.15, ${\rm K}_{\rm s}$=4.548), Gliese 687 has been heavily studied, and in particular, the CHARA Array has recently been used to obtain direct interferometric angular diameter measurements for the star. \citet{Boyajian12} find $R_{\star}/R_{\odot}=0.4183\pm0.0070$, and derive $L_{\star}/L_{\odot}=0.02128\pm0.00023$, $T_{\rm eff}=3413\,{\rm K}$, and use the mass-radius relation of \citet{HM93} to obtain $M_{\star}/M_{\odot}=0.413\pm0.041$. As illustrated in Figure \ref{fig:sindex}, Gliese 687's mean Mt. Wilson S value and the dispersion of its S-index measurements from our spectra indicate that it has a moderate degree of chromospheric activity. This conclusion is in concordance with our long-term photometric monitoring program, which also indicates that the star is somewhat active.

 \begin{deluxetable}{ccc}
 \tablecaption{Stellar Parameters for Gliese 687
 \label{tab:stellarparams}}
 \tablecolumns{3}
 \tablehead{{Parameter} & {Value} & {Reference}}
 \startdata
 Spectral Type & M3 V & \citep{Rojas-Ayala12} \\
 Mass ($M_{\odot}$) & $0.413\pm0.041$ & \citep{Boyajian12} \\
 Radius ($R_{\odot}$) & $0.4183\pm0.0070$ & \citep{Boyajian12} \\
 Luminosity ($L_{\odot}$) & $0.0213\pm0.00023$ & \citep{Boyajian12} \\
 Distance (pc) & $4.5\pm0.115$ & \citep{Rojas-Ayala12} \\
 $B-V$ & 1.5 & Simbad\\
 $V$ Mag. & 9.15 & \citep{Rojas-Ayala12}\\
 $J$ Mag. & 5.335 & \citep{Cutri03}\\
 $H$ Mag. & 4.77 & \citep{Cutri03}\\
 $K$ Mag. & 4.548 & \citep{Cutri03}\\
 Avg. S-index & 0.811 & This work \\
 $\sigma_{\rm S-index}$ & 0.096 & This work \\
 $P_{\rm rot}$ (days) & $61.8\pm 1.0$ & This work \\
 $T_{\rm eff}$ (K) & $3413\pm28$ & \citep{Boyajian12} \\
 \enddata
 \end{deluxetable}

 \section{Radial Velocity Observations}

 Doppler shifts from both the Keck (122 observations) and APF (20 observations) platforms were measured, in each case, by placing an iodine absorption cell just ahead of the spectrometer slit in the converging beam of stellar light from the telescope \citep{Butler96}. The forest of iodine lines superimposed on the stellar spectra generates a wavelength calibration and enables measurement of each spectrometer's point spread function. The radial velocities from Keck were obtained by operating HIRES at a spectral resolving power R$\sim$70,000 over the wavelength range of 3700-8000 $\AA$, though only the region 5000-6200 $\AA$ containing a significant density of iodine lines was used in the present Doppler analysis \citep{Vogt94}. The APF measurements were obtained over a similar spectral range, but at a higher spectral resolving power, R$\sim$108,000. For each spectrum that was obtained, the region containing the iodine lines was divided into $\sim$700 chunks, each of $\sim2\,\AA$ width. Each chunk produces an independent measure of the wavelength, PSF, and Doppler shift. The final measured velocity is the weighted mean of the velocities of the individual chunks. All radial velocities (RVs) have been corrected to the solar system barycenter, but are not tied to any absolute velocity system. As such, they are ``relative'' velocities, with a zero point that can float as a free parameter within an overall system model.

 The internal uncertainties quoted for all the radial velocity measurements in this paper reflect only one term in the overall error budget, and result from a host of errors that stem from the characterization and determination of the point spread function, detector imperfections, optical aberrations, consequences of undersampling the iodine lines, and other effects. Two additional major sources of error are photon statistics and stellar ``jitter''. The latter varies widely from star to star, and can be mitigated to some degree by selecting magnetically inactive older stars and by time-averaging over the star's unresolved low-degree surface $p$-modes. All observations in this paper have been binned on 2-hour timescales. In addition to the radial velocities that we have obtained at Keck and APF, we also use five Doppler measurements obtained by \citet{Endl2003} at the Hobby-Eberly Telescope located at McDonald Observatory. These radial velocity observations are presented in the appendix.

 \section{The Best Fit Solution}

 The combined radial velocity data sets show a root-mean-square (RMS) scatter of 7.58 ${\rm m\,s}^{-1}$ about the mean velocity. This scatter is measured after we have applied best-fit telescope offsets of 0.64 ${\rm m\,s^{-1}}$ for Keck, -1.71 ${\rm m\,s^{-1}}$ for APF, and 1.27 ${\rm m\,s^{-1}}$ for HET.

\begin{figure}
 \plotone{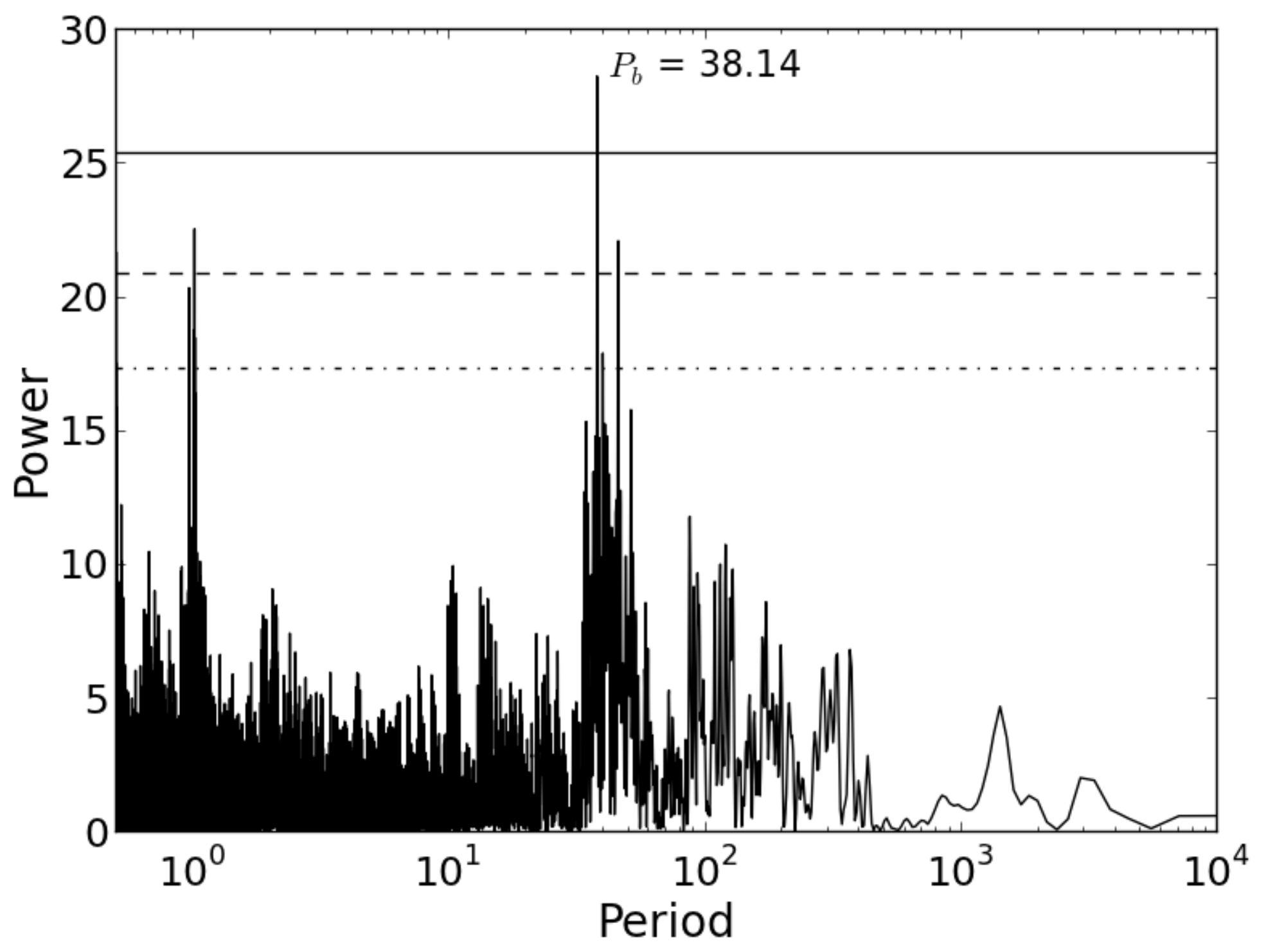}
 \caption{Lomb-Scargle periodograms for combined radial velocity measurements of GJ~687 from the HET, Keck and APF telescopes. The horizontal lines from top to bottom represent false alarm probabilities of 0.01\%, 0.1\%, and 1.0\% respectively.}
 \label{fig:periodogram}
 \end{figure}

 A Lomb-Scargle periodogram of the 149 velocity measurements of GJ~687 is shown in Figure \ref{fig:periodogram}. False alarm probabilities are calculated with the bootstrap method, as described in \citet{Efron79}, iterating 100,000 times for a minimum probability of ${\rm P}_{\rm false} < 1e-5$ as easily met by the tallest ${\rm P}_{{\rm b}}$=38.14 day peak in Figure \ref{fig:periodogram}. This signal in the data is modeled as a $M_{\rm b}\sin(i)=0.06\,M_J$ planet with an orbital eccentricity, $e_b=0.04$.

 \begin{deluxetable}{lllr}
 \tablecaption{1-planet model for the GJ~687 System\label{tab:fit}}
 \tablecolumns{3}
 \tablehead{{Parameter}&{Best fit}&{Errors}}
 \startdata
 Period (d) & 38.14&(0.015) \\
Mass ($M_{J}$) & 0.058&(0.007) \\
Mass ($M_{\oplus}$) & 18.394&(2.167) \\
Mean Anomaly (deg) & 234.62&(87.962) \\
Eccentricity & 0.04&(0.076) \\
Longitude of periastron (deg) & 359.43&(120.543) \\
Semi-major Axis (AU) & 0.16353&(0.000043) \\
& & & \\
Time of Periastron (JD) & 2450579.11&(9.32) \\
& & & \\
RV Half Amplitude (${\rm m\,s^{-1}}$) & 6.43&(0.769) \\

 &&\\
 First Observation Epoch (JD)& 2450603.97 &\\
 Velocity Offsets &&\\
 ~~Keck/HIRES & 0.64 $\,{\rm m\,s^{-1}}$ & (0.63) \\
 ~~APF/Levy & -1.71 $\,{\rm m\,s^{-1}}$ & (1.68) \\
 ~~HET & 1.27 $\,{\rm m\,s^{-1}}$ & (0.98) \\
 $\chi^2$ & 18.55 &\\
 RMS && \\
 ~~Keck/HIRES & 6.62 $\,{\rm m\,s^{-1}}$ &\\
 ~~APF/Levy & 3.95 $\,{\rm m\,s^{-1}}$ &\\
 ~~HET & 2.44 $\,{\rm m\,s^{-1}}$ &\\
 Jitter & 5.93 $\,{\rm m\,s^{-1}}$ &\\
 \enddata
 \tablecomments{All elements are defined at epoch JD = 2450603.97. Uncertainties are reported in parentheses.}
 \end{deluxetable}

\begin{figure}
 \plotone{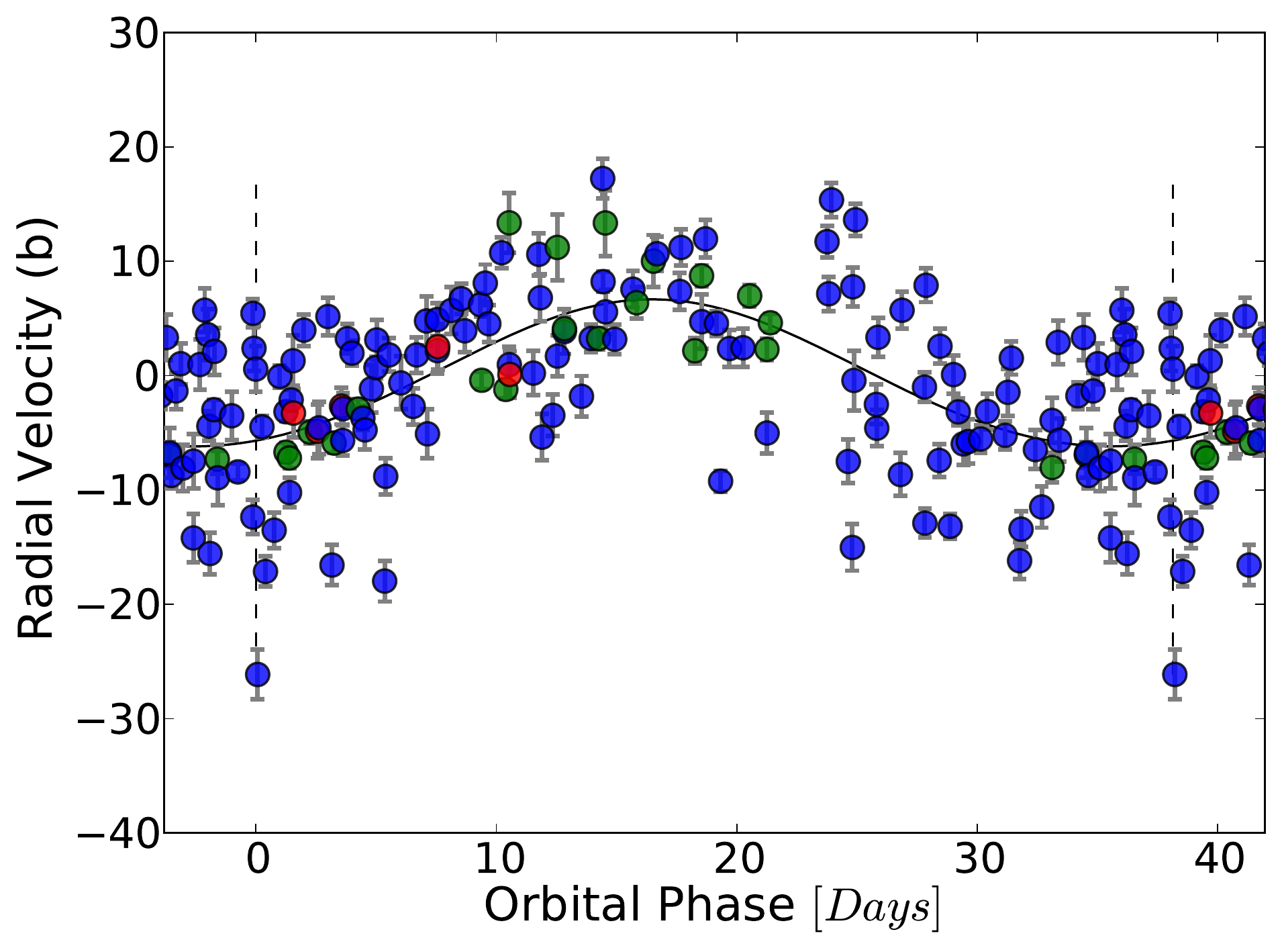}
 \caption{Phased radial velocity model for planet b, folded at the $P=38.14\,{\rm d}$ orbital period. The blue points correspond to Keck data points, green points to APF data, and the red points are HET data. The vertical dashed lines demarcate the extent of unique data.}
 \label{fig:phasedCurves}
 \end{figure}

\begin{figure}
 \plotone{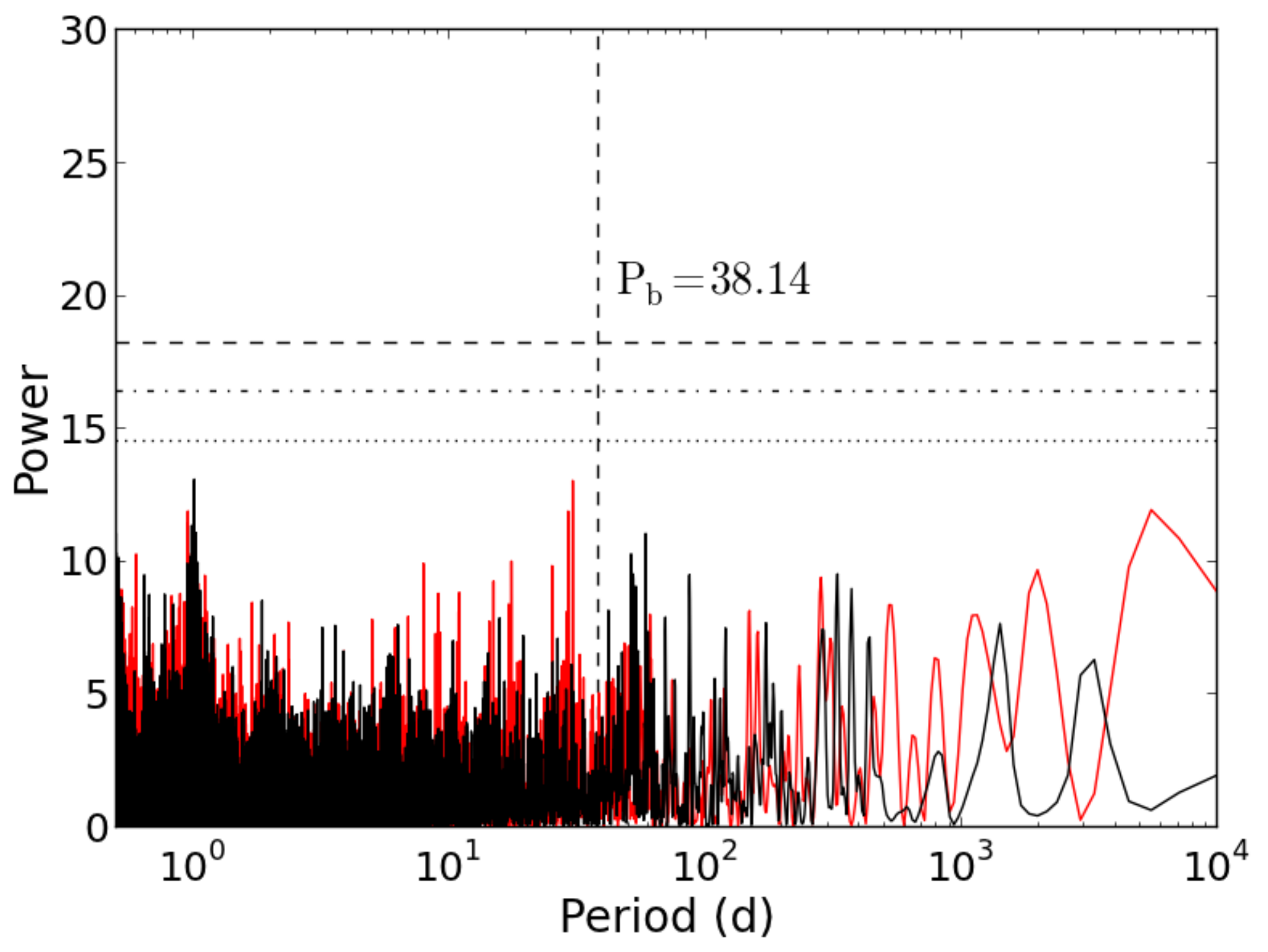}
 \caption{Lomb-Scargle periodogram of the radial velocity residuals to the fit given in Table \ref{tab:fit} plotted in black, and the Lomb-Scargle periodogram of the Mt. Wilson S-index values plotted behind in red.}
 \label{fig:residPeriodogram}
 \end{figure}

 Using Levenberg-Marquardt optimization, we obtained a best-fit Keplerian model for the system. This fit, which assumes $i=90^{\circ}$ and $\Omega=0^{\circ}$ for the planet, is listed in Table \ref{tab:fit}. The phased RV curve for the planet in Table \ref{tab:fit} is shown in Figure \ref{fig:phasedCurves}. A power spectrum of the residuals to our one-planet fit is shown in Figure \ref{fig:residPeriodogram} and indicates no significant periodicities. Also shown in this figure is a periodogram of our Mt. Wilson S-index measurements from the spectra, which are a proxy for the degree of spot activity on the star at a given moment. None of the peaks in the periodogram of S-index values coincide with the peak that we suspect to be a planet.

 \begin{figure}
 \plotone{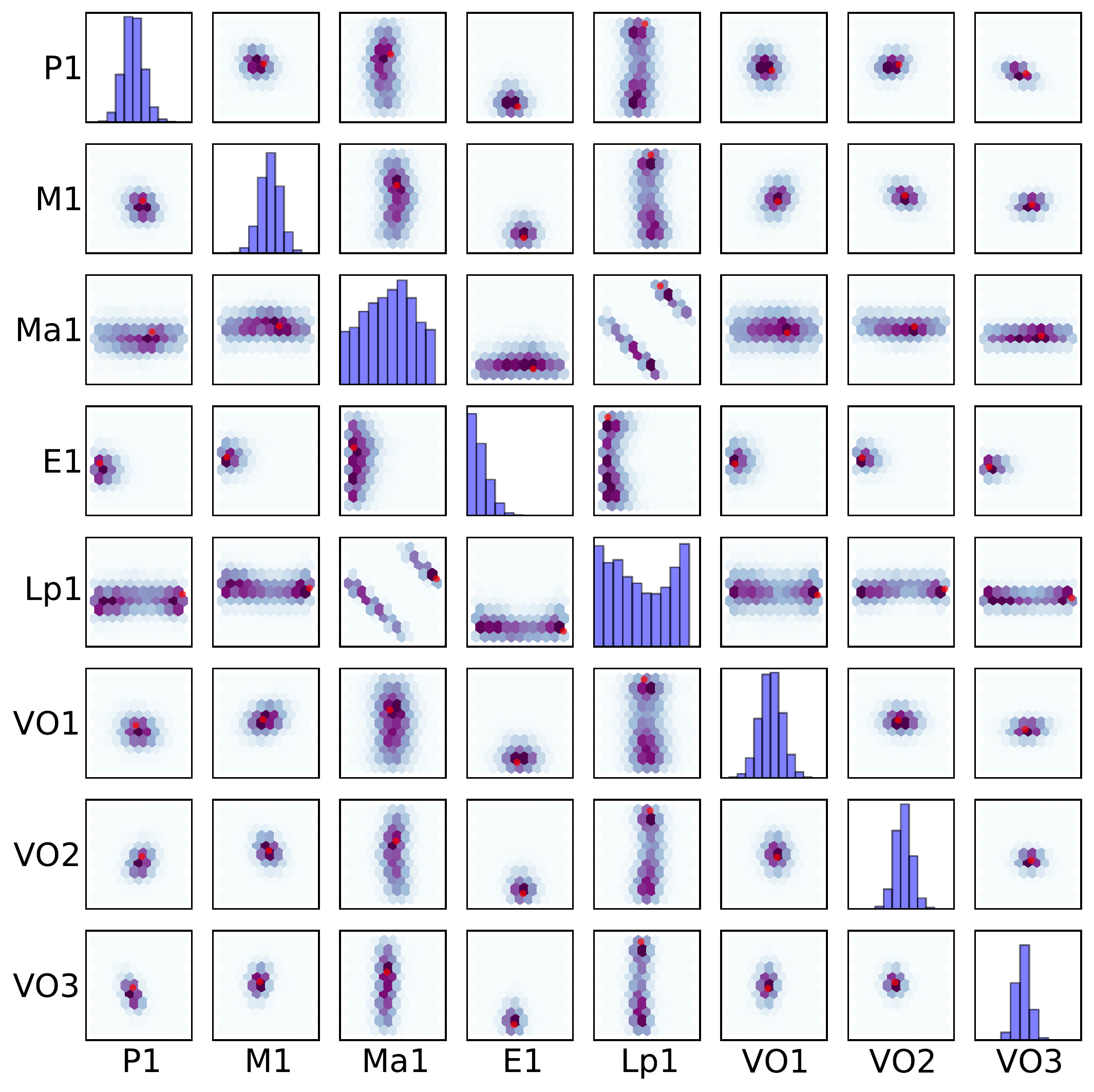}
 \caption{Smooth scatter plots of parameter error correlations for our Markov chain. In each case, the best-fit model is indicated with a small red dot, and the density of models within the converged portion of the chain is shown as a blue-toned probability distribution function. The diagonal line of entries shows the marginalized distribution for each parameter of the one-planet model.}
 \label{fig:errorCorr}
 \end{figure}

 \begin{figure}
 \plotone{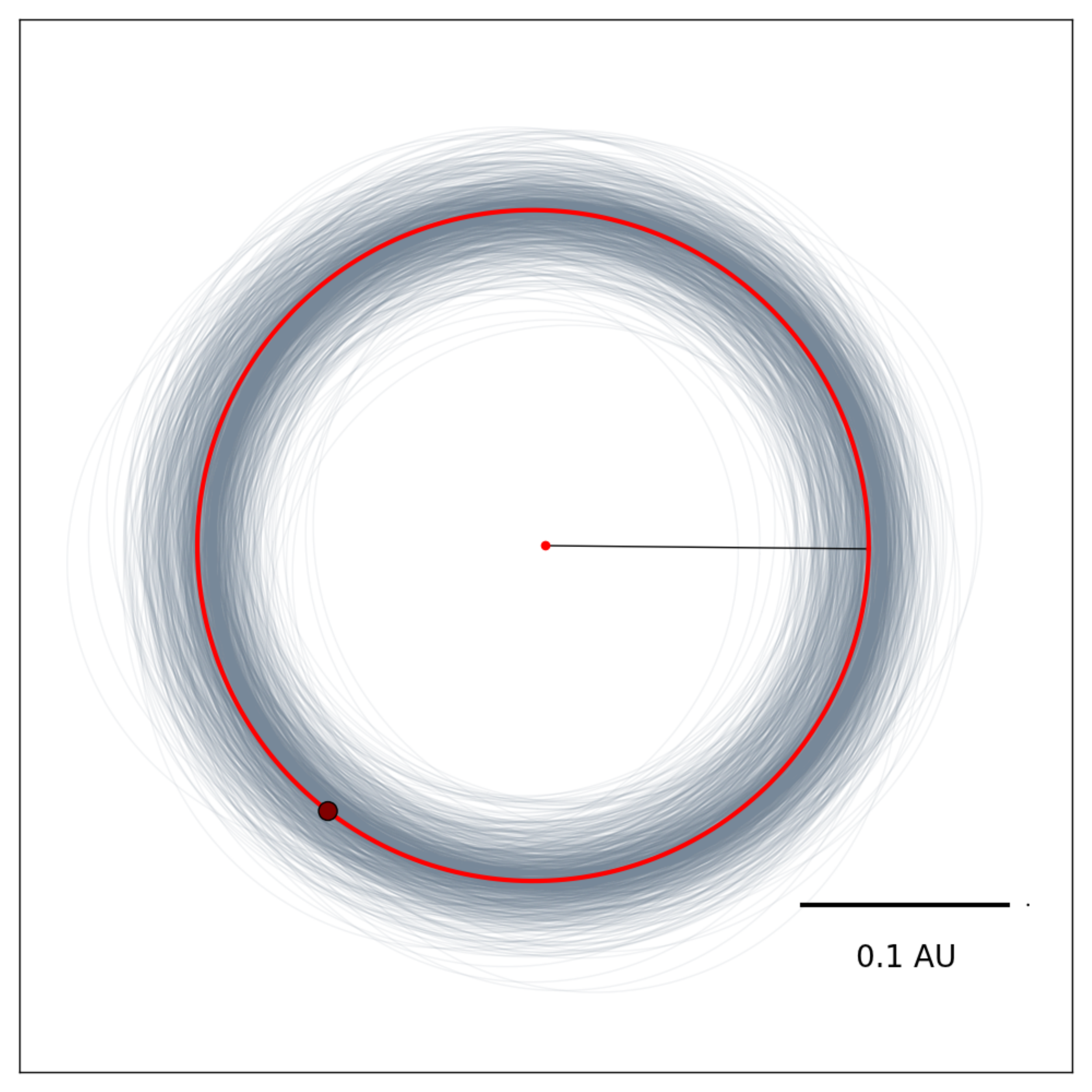}
 \caption{The orbit of the proposed planetary companion to GJ~687. The larger red point corresponds to the location of the planet at the initial observation epoch, HJD 2450603.97. The line from the origin corresponds to the planet's periastron. For the geometry plotted, transits, should they occur, would happend when the planet traverses the positive y-axis. The light lines are 100 orbits of the planet drawn from the converged segment of the Markov Chain. The red dot in the center of the diagram corresponds to the size of the star when drawn to scale. The small black dot next to the distance scale bar indicates the size of the planet when drawn to scale, and assuming it has $R_{\rm P} = R_{\rm Nep}$.}
 \label{fig:orbit}
 \end{figure}

 The reduced chi-squared statistic for our fit is $\chi_{red}^2 =$ 18.55 and results in a fit with a combined RMS of 6.16 ${\rm m\,s^{-1}}$ and estimated excess variance of $\sigma_{\rm jitter}$ = 5.93 ${\rm m\,s^{-1}}$ (the estimate of the jitter that is required to bring the reduced chi-squared statistic of the fit down to unity). This value accounts for variance in both the stellar signal and from the telescope itself, though for a moderately active star such as GJ~687 a stellar jitter of order 6.0 ${\rm m\,s^{-1}}$ is reasonable,
 and could account for the majority of the observed variance.

 In order to compute parameter uncertainties for our orbital fit, we implement a Markov Chain Monte Carlo algorithm \citep{Ford05, Ford06, Balan09, Meschiari09, Gregory11}. The MCMC algorithm returns a chain of state vectors, ${\bf k}_{i}$ (a set of coupled orbital elements, e.g. period, mass, etc. and the three velocity offset parameters). The goal of the Markov Chain calculation is to generate an equilibrium distribution proportional to $\exp[\chi^{2}({\bf k})]$. We adopt non-informative priors on all parameters (and uniform in the log for masses and periods). The resulting error correlations are shown in Figure \ref{fig:errorCorr}, and a set of 100 states drawn randomly from the converged chain are shown in the orbital diagram in Figure \ref{fig:orbit}.

\begin{figure}
 \plotone{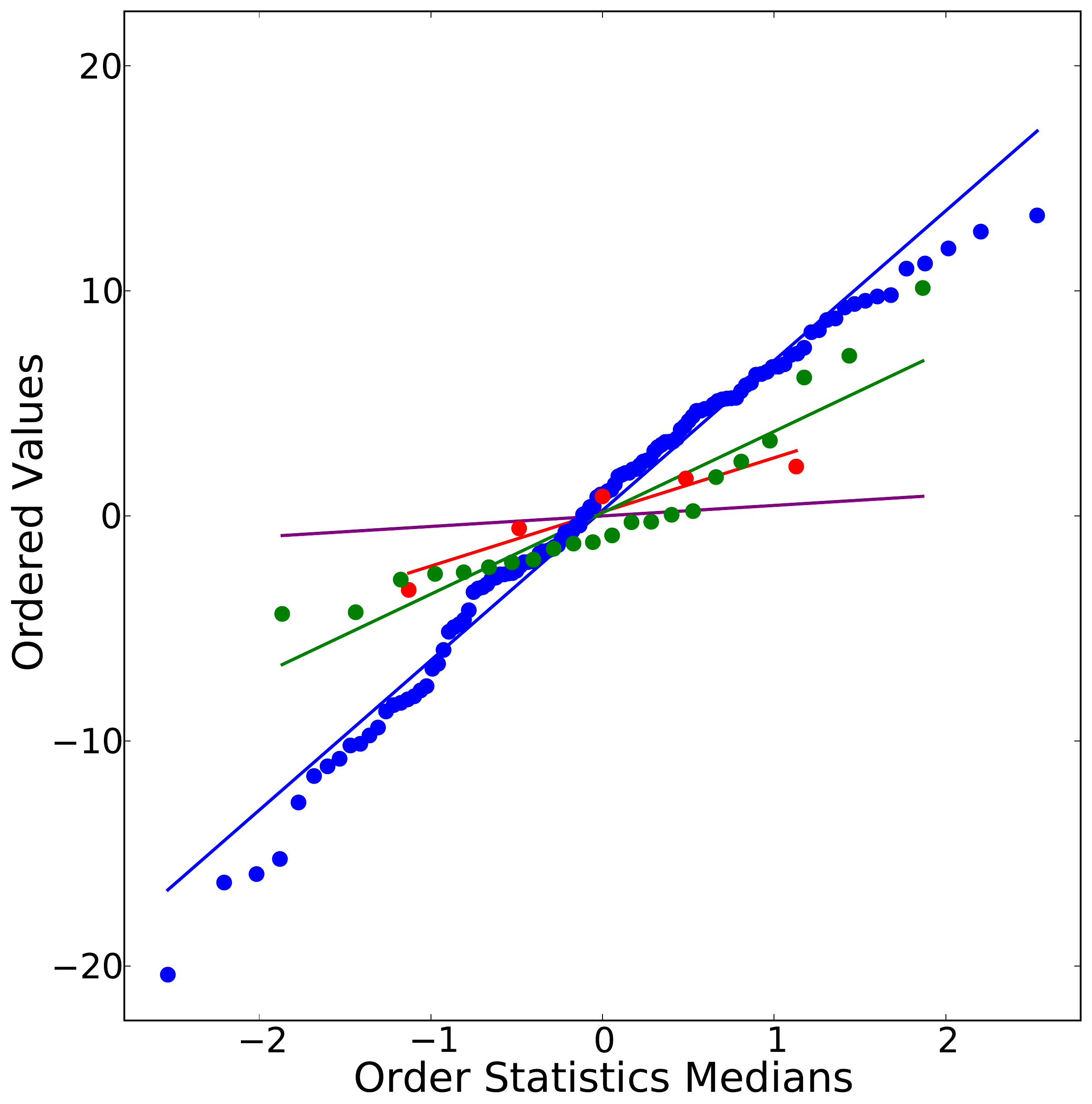}
 \caption{Quantile-Quantile plot for the velocity residuals to the 1-planet model fit. Adherence of the points to the lines indicate the degree to which the radial velocities from the two telescopes conform to a normal distribution. APF points are shown in green, Keck points are shown in blue, and HET points are shown in red.}
 \label{fig:fitPlot}
 \end{figure}

 The error correlation diagram indicates that all parameters are well determined, save the usual degeneracies between mean anomaly and $\omega$ for the low-eccentricity orbit. The distribution of the residuals relative to the best-fit model shows no evident pathologies. Indeed, a quantile-quantile plot (shown in Figure \ref{fig:fitPlot}) indicates that the distribution of residuals is well described by a normal distribution. We note that the smaller scatter of points obtained with the APF telescope could be a consequence of the fact that they were all taken within a $\Delta t$ = 140d period, and thus sample only one segment of the stellar activity cycle.

\begin{figure}
 \plotone{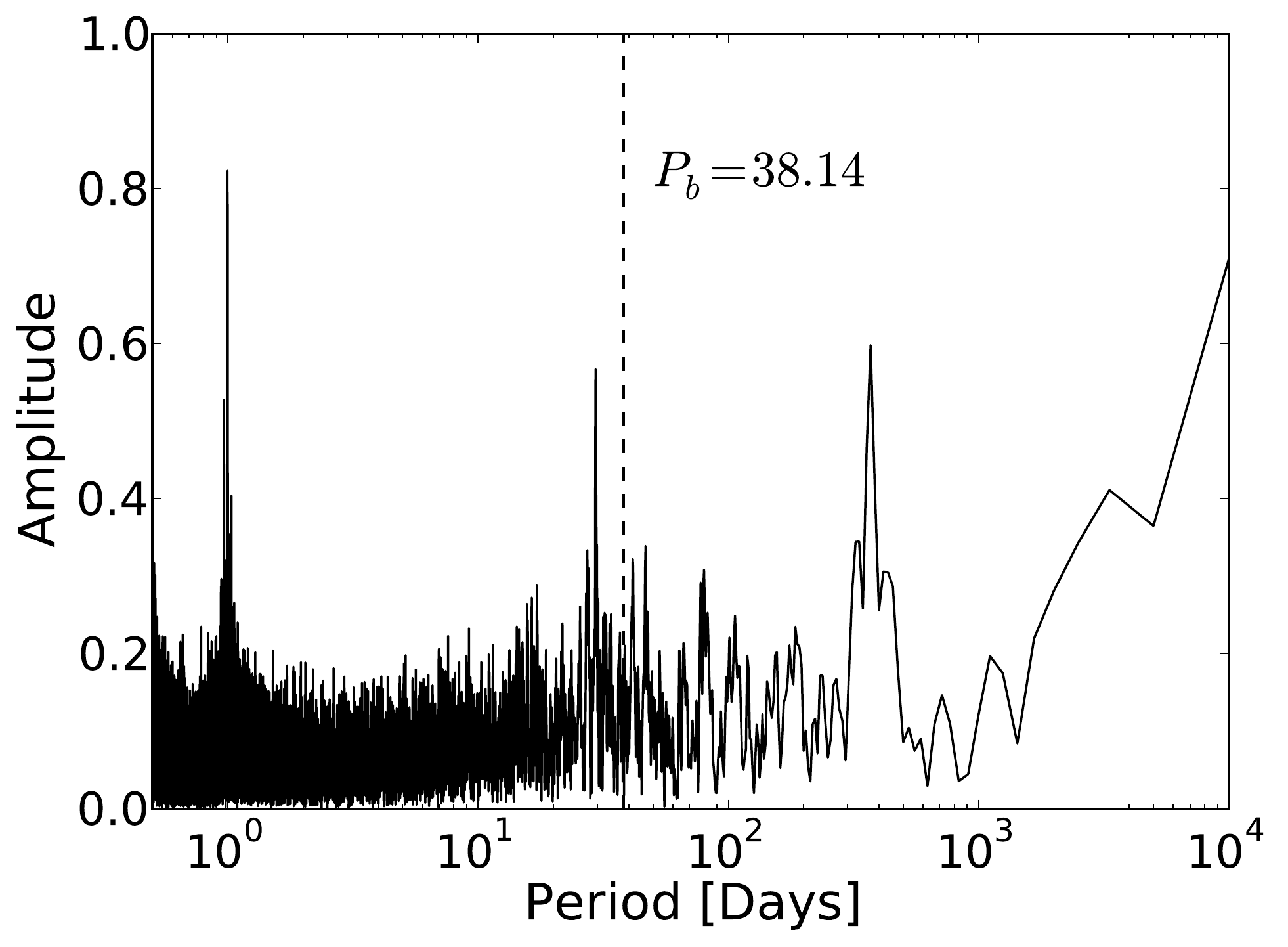}
 \caption{Window function calculated from all radial velocity observations of GL~687. While several peaks exist due to aliasing effects from our data's time stamps, none of them coincide with the locations necessary to create a peak in the periodogram at our best fit period of $P=38.14$ days}
 \label{fig:windowFunc}
 \end{figure}

 A potentially significant challenge to correctly identifying the orbital period of a
 proposed exoplanet arises from the discrete and uneven sampling inherent
 in radial velocity surveys. The spacing of observations leads to increased noise and the
 presence of aliases within the star's periodogram which can be mistaken for
 a true orbital signature. For a real signal occurring at a frequency $f_{planet}$
 we expect alias signatures at $f=f_{planet} \pm n f_{sampling}$ where n is an integer. In order to aid confirmation that the periodic signal we observe is actually a planetary signature, we must be able to calculate where aliases due to our observing cadence will occur, and
 then verify that they are not the source of the signal. The aliases are determined using a
 spectral window function as defined by \citet{Roberts87}

 \begin{equation}
 W(\nu) = \frac{1}{N}\sum\limits_{r=0}^N \exp^{-2\pi{\it i}\nu t_r},
 \end{equation}

 \noindent where N is the total number of observations and t is the date on which they were taken. Plotting
 this function will result in peaks that are due solely to the sampling cadence of the data. Because our observations are constrained by when the star is visible in the night sky, and because Keck Telescope time is allocated to Doppler surveys primarily when the Moon is up, we expect aliases at periods of 1 solar day, 1 sidereal day, 1 synodic month and 1 sidereal year. Examining the window function in Figure \ref{fig:windowFunc} we do see peaks resulting at these periods, but careful analysis of the periodogram for our radial velocity observations shows no evidence of strong signals occurring at the locations necessary for our $P=38.14$ day signal to be a potential alias instead of a true Keplerian signature.

 With an apparent ${\rm K}_{\rm s}$-band magnitude of 4.54, Gliese 687 is brighter (in the near infrared) than {\it all} known hosts of transiting extrasolar planets other than 55 Cancri. As a consequence, transits by Gliese 687's planetary companion (which has an equilibrium temperature, $T_{\rm eq}\sim260\,{\rm K}$), were they to occur, would be of substantial scientific value. In particular, transmission spectroscopy with JWST would give insights into what is likely a dynamic and chemically rich planetary atmosphere. The {\it a-priori} geometric transit probability for Gliese 687~b, however, is a scant $P_{\rm tr}=1.2$\%, and as we describe below, there is no evidence that transits occur. With $M\sin(i)=19\,M_{\oplus}$, the currently observed mass-radius range for exoplanets indicates that the planetary radius, $R_{\rm p}$ could credibly range from $R_{\rm p}\sim0.2R_{\rm Jup}$ to $R_{\rm p}\sim0.6R_{\rm Jup}$, implying potential transit depths in the $d=0.2$\% to $d=2$\% range.

 \section{Photometric Observations}

 During the 2009--2013 observing seasons, we acquired a total of 866
 photometric observations of GJ~687 on 519 nights with the Tennessee State
 University (TSU) T12 0.80~m automatic photoelectric telescope (APT) at
 Fairborn Observatory in Arizona. The T12 APT is one of several TSU
 automatic telescopes operated at Fairborn \citep{Henry99,Eaton2003}. It is
 equipped with a two-channel precision photometer that employs a dichroic
 filter and standard Str\"omgren $b$ and $y$ filters to separate the two
 passbands and two EMI 9124QB bi-alkali photomultiplier tubes to measure
 the $b$ and $y$ count rates simultaneously. We observed GJ~687, designated
 our program star (P), differentially with respect to three neighboring
 comparison stars: C1 (HD~156295, $V=5.54$, $B-V=0.22$, F0~IV), C2
 (HD~160198, $V=7.65$, $B-V=0.46$, F2~V), and C3 (HD~161538, $V=7.01$,
 $B-V=0.44$, F2~V). A detailed description of the observing sequence and
 the data reduction and calibration procedures are given in \citet{Henry99}.

 We computed all pairwise differential magnitudes $P-C1$, $P-C2$, $P-C3$,
 $C3-C2$, $C3-C1$ and $C2-C1$ in both the $b$ and $y$ passbands, corrected
 them for atmospheric extinction, and transformed them to the standard
 Str\"omgren photometric system. Observations with internal standard
 deviations greater than 0.01 mag were discarded to remove data taken in
 non-photometric conditions. Intercomparison of the six sets of differential
 magnitudes demonstrated that HD~156295 (C1) is a low-amplitude variable
 while both HD~160198 (C2) and HD~161538 (C3) are constant to the expected
 measurement precision. To improve our precision, we combined the separate
 differential $b$ and $y$ observations into a single $(b+y)/2$ ``passband."
 We also computed the differential magnitudes of GJ~687 with respect to the
 mean brightness of the two good comparison stars: $P-(C2+C3)/2$. The
 standard deviation of the $C3-C2$ comparison star differential magnitudes
 is 0.0020 mag, which we take to be the precision of a single measurement.

 \begin{figure}
 \plotone{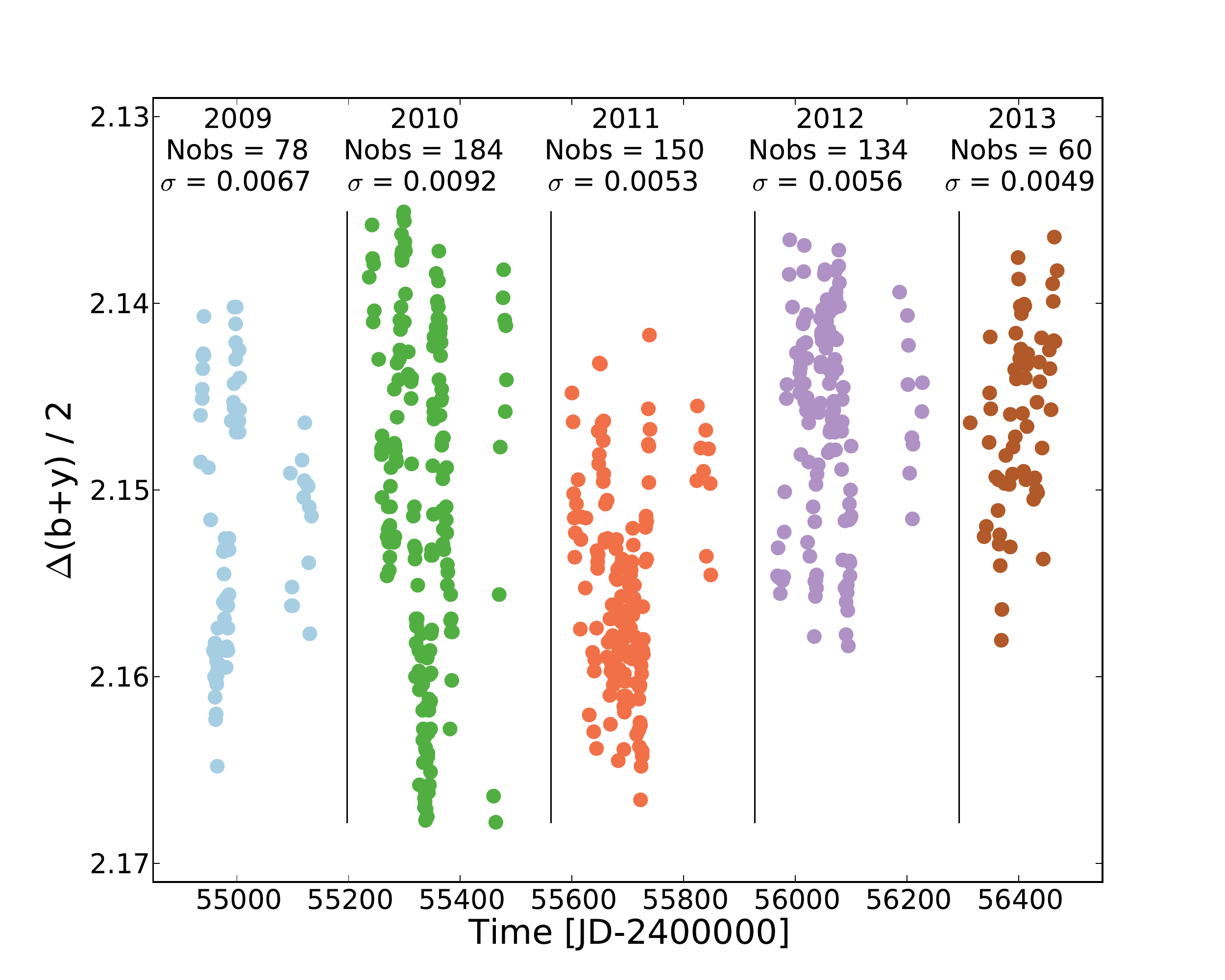}
 \plotone{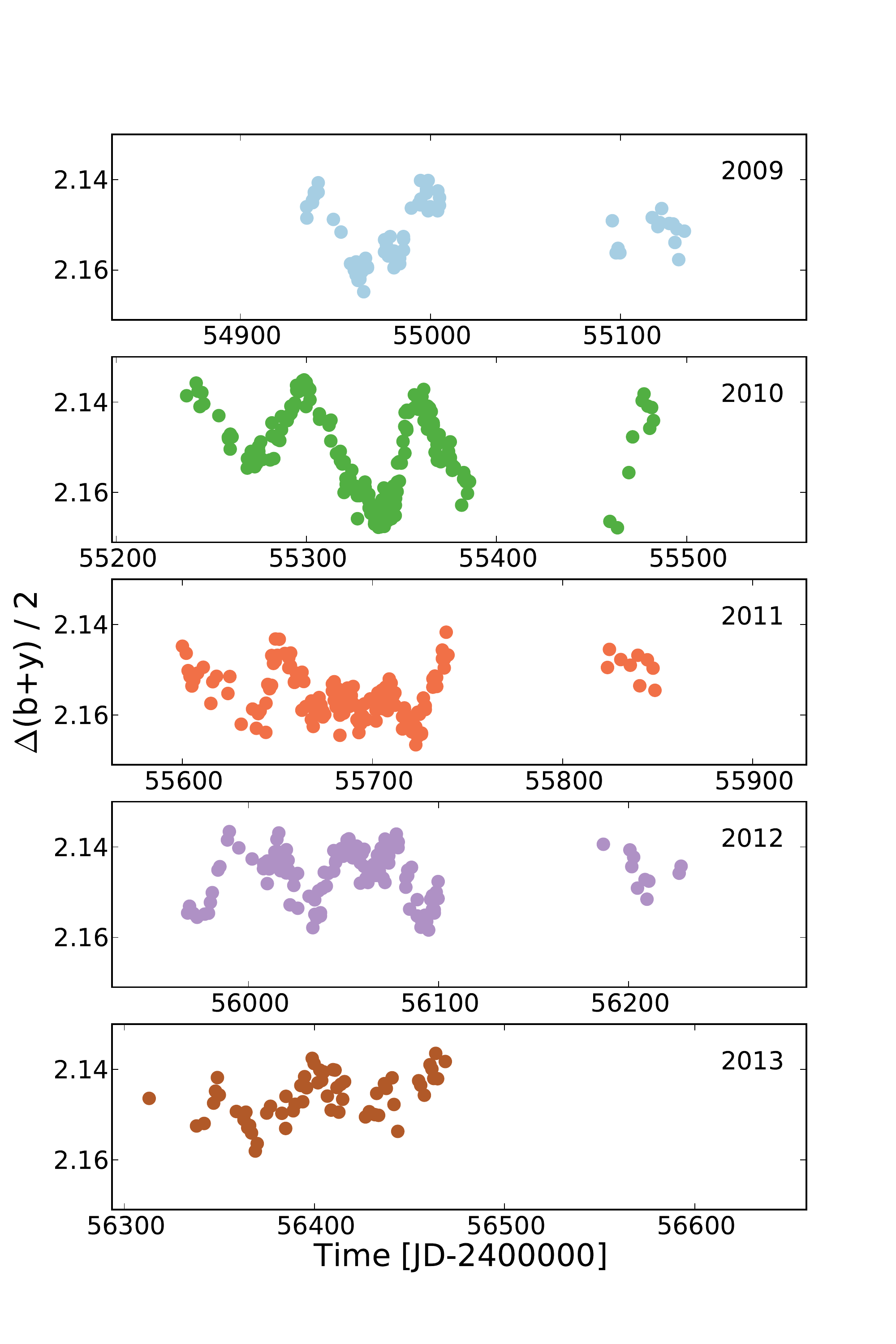}
 \caption{ Photometric data taken of GJ~687 over 5 years. The top panel shows the total data set with information regarding observations and standard deviation for each year. The bottom panel gives a closer look at the data seperated by year.}
 \label{fig:phot1}
 \end{figure}

 \begin{figure}
 \plotone{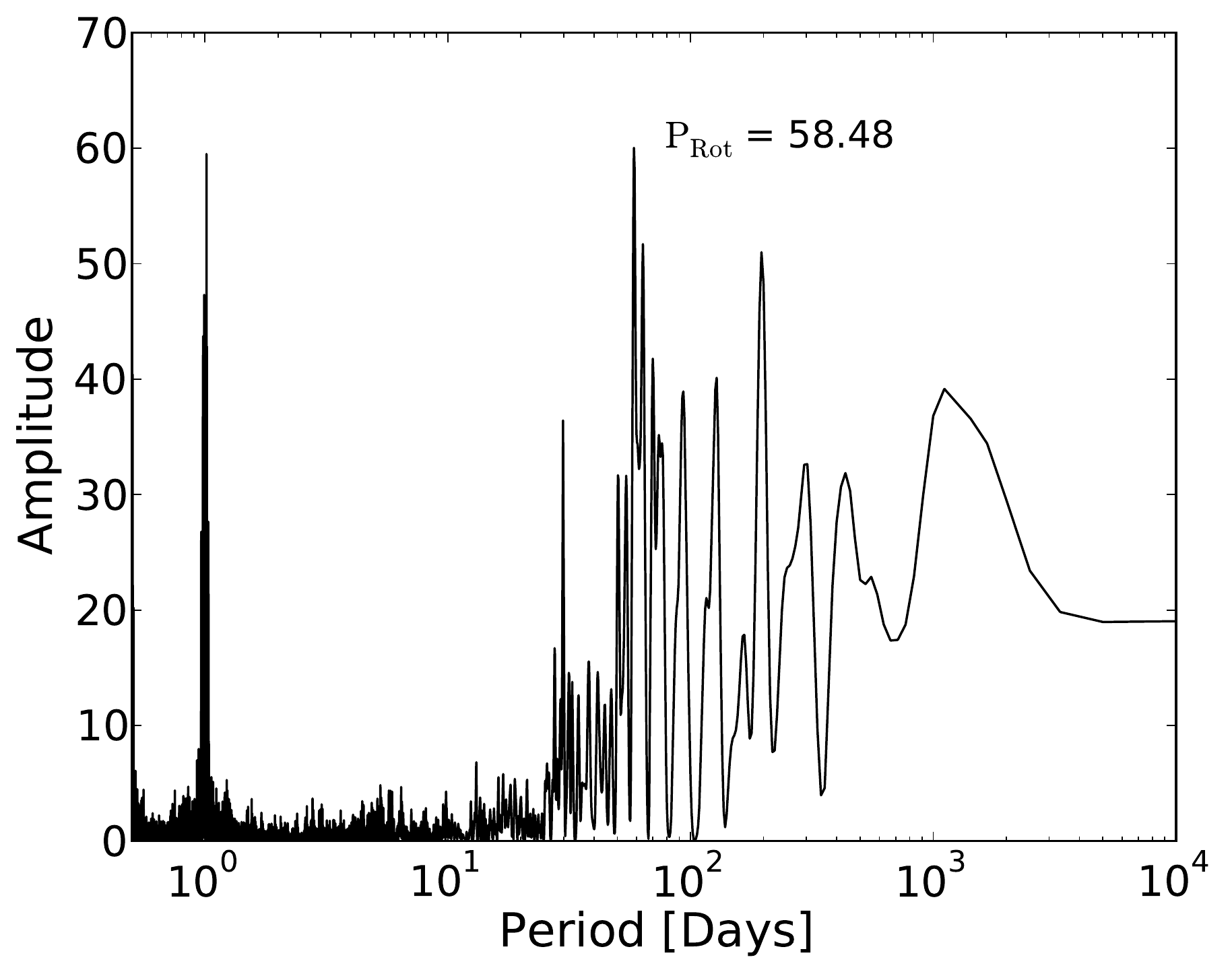}
 \plotone{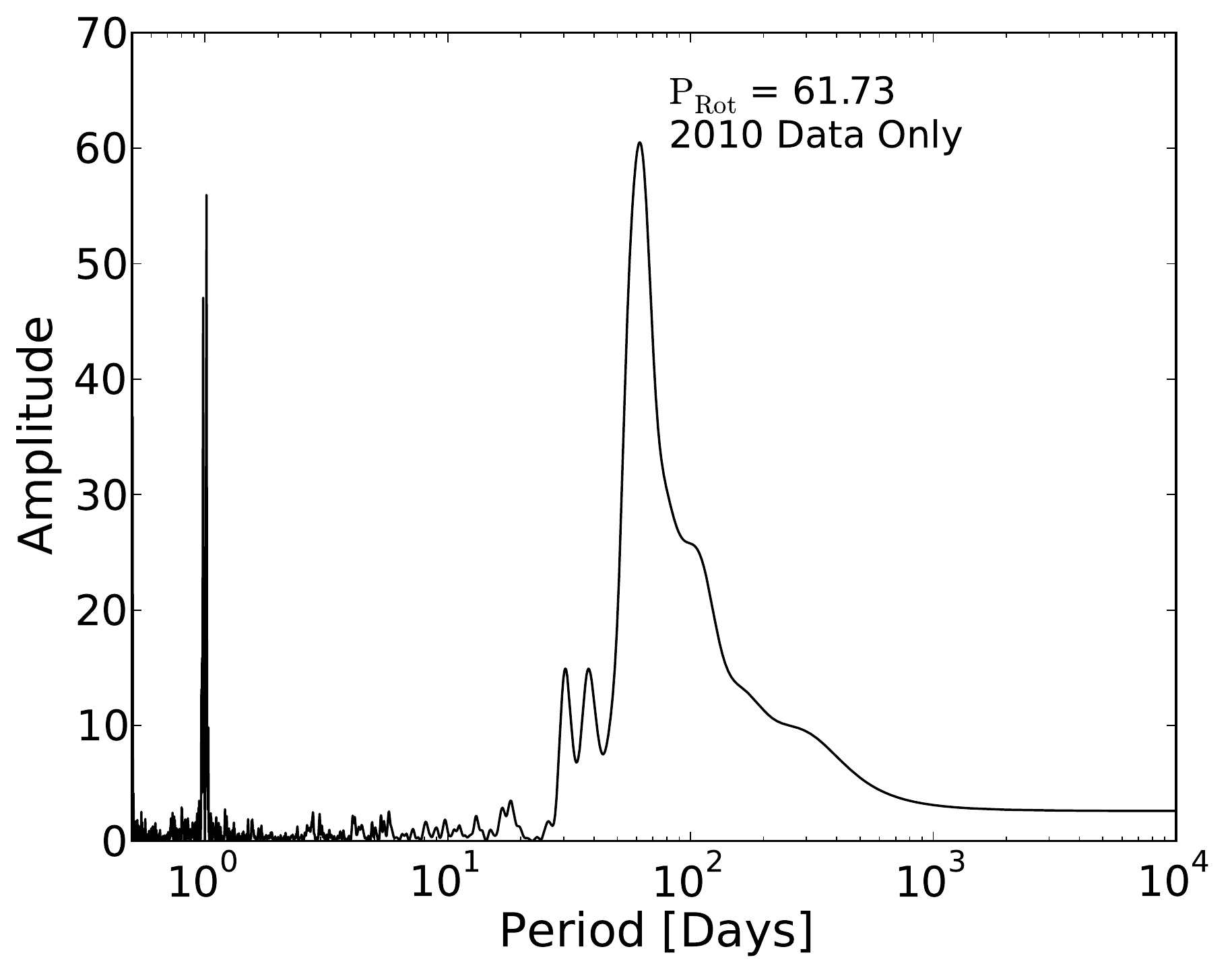}
 \caption{ Lomb-Scargle periodogram of the photometric observations of GJ~687. In the combined data set, the maximum observed power occurs at 58.48 days (top panel). However when we consider only data obtained in 2010, where the rotational modulation is most clearly exhibited, we find that maximum power occurs at ${\rm P} = 61.73$ days (bottom panel). We identify this periodicity with the rotational period of the star.}
 \label{fig:phot2}
 \end{figure}

 \begin{figure}
 \plotone{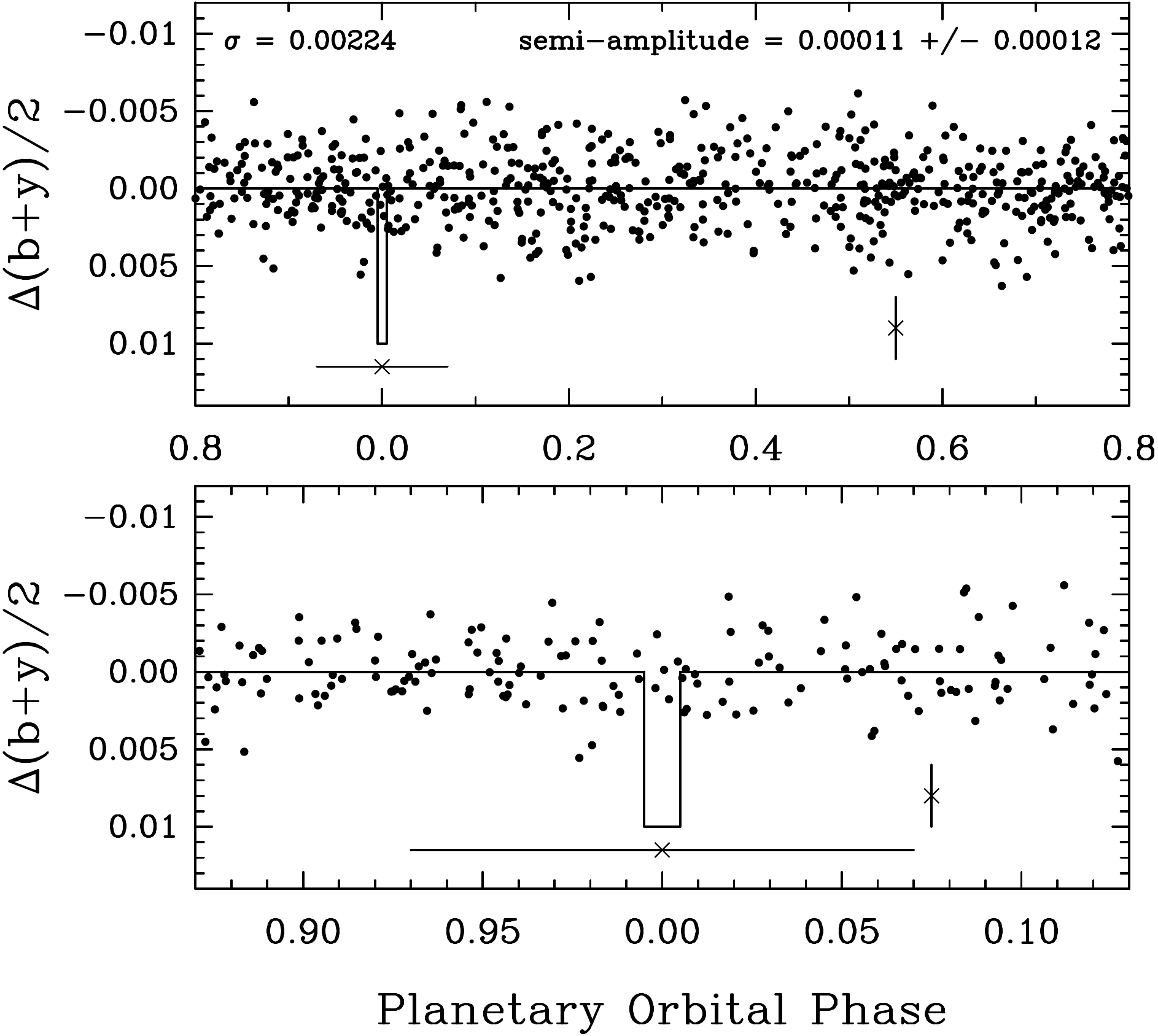}
 \caption{ {\it Top panel} Filtered differential photometric measurements for Gliese~687 folded at the best-fit planetary period, ${\rm P}=38.14$ days. A light curve model for a centrally transiting Neptune-sized planet is shown. The vertical error bar indicates the 0.002 magnitude photometric precision. The horizontal error bar shows the 1-$\sigma$ uncertainty on the time of a central transit. {\it Bottom panel} shows a magnified view of the folded photometric data in the vicinity of the predicted time of central transit.}
 \label{fig:phot3}
 \end{figure}

 \begin{figure}
 \plotone{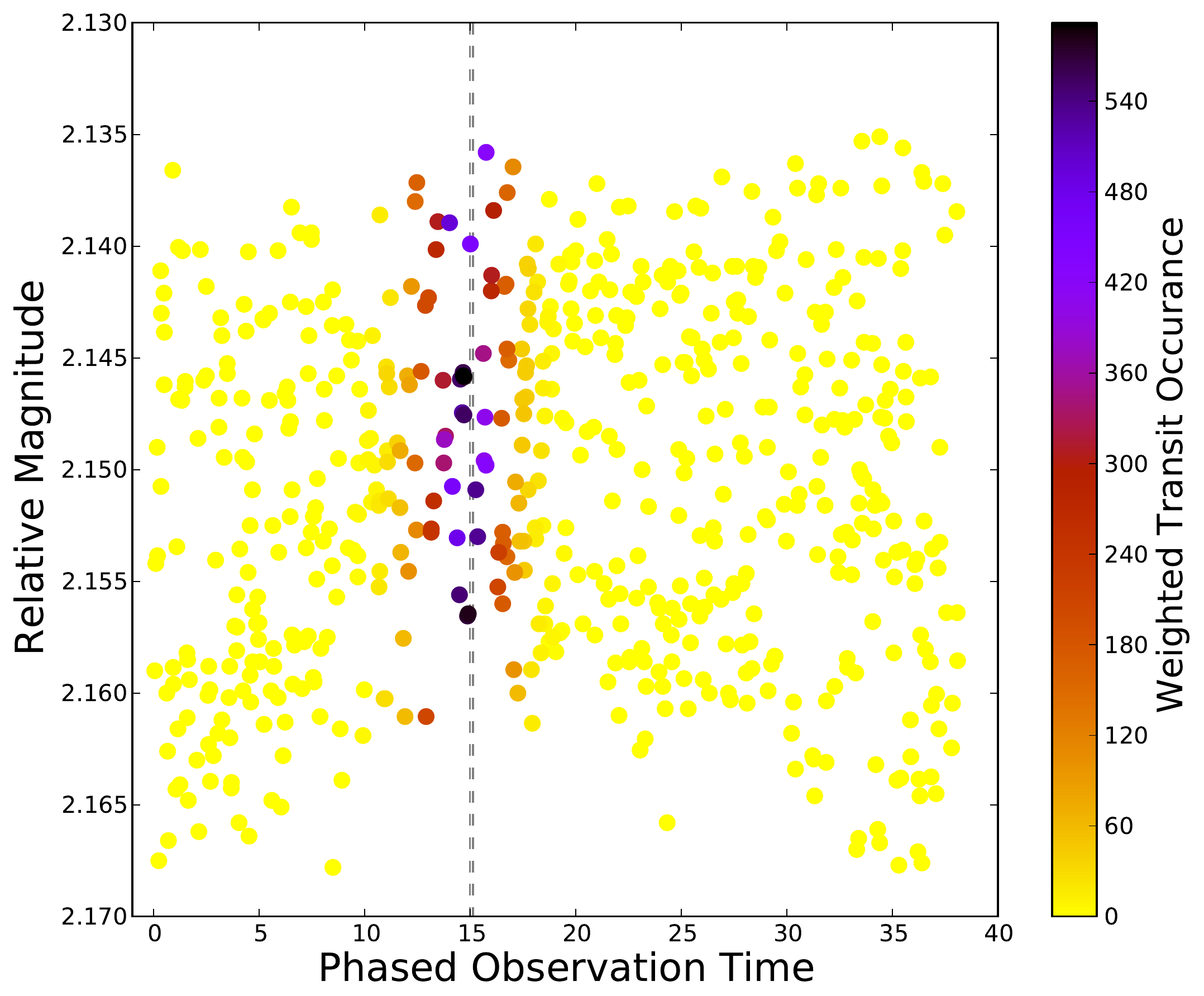}
 \caption{For each of the 10,402 potential systems in our Markov chain, we check the predicted transit times against our photometric observations. If a photometric data point lies within the transit window of a particular member of the Markov chain, we assign a value to that point which is cosine-weighted by its distance from the predicted time of central transit. The sum of these values is mapped onto the color of the points in the diagram. The phase of the points, as well as the vertical gray bar spanning the predicted 3-hour central transit duration are for our best fit model given in Table \ref{tab:fit}.}
 \label{fig:transit}
 \end{figure}

 A total of 606 nightly measurements in the five observing seasons survived
 the cloud-filtering process. These data are plotted as $P-(C2+C3)/2$
 differential magnitudes in the top panel of Figure \ref{fig:phot1}. The five individual
 observing seasons are plotted in the remaining panels. The standard
 deviations for the yearly light curves are given in each panel. These
 range from 0.0049 to 0.0092 mag, compared to the measurement precision of
 0.0020 mag. Gaps of 10--12 weeks in the yearly light curves for 2009 through 2012 are due to southern Arizona's July--September rainy season when good photometry is
 not possible.

 Low-amplitude variability is seen in GJ~687 during each observing season,
 resembling light curves typical of modestly active stars with spot filling
 factors of a few percent \citep[see, e.g.,][]{Henry95}. The 2010 light
 curve has the largest amplitude variability ($\sim0.03$~mag) and reveals cyclic
 variation with a time scale of $\sim60$~days. The other light curves
 have lower amplitudes and include cyclic variations of $\sim60$ and also
 $\sim30$ days. These year-to-year and cycle-to-cycle variations are also
 typical of modestly active stars. We interpret the 60-day variability as
 the signature of the star's rotation period and the 30-day variability as
 a sign of spot activity on opposite hemispheres of the star.

 Frequency spectra of the complete 2009 - 2013 data set and of the 2010 data alone are
 shown in the top and bottom panels of Figure \ref{fig:phot2} respectively. The rotational
 modulation signal is seen most clearly in the 2010 data, which matches up with the most
 coherent light curve in Figure \ref{fig:phot1}. Therefore, we take the $58.48 \pm 1.0$d signal as
 our best measurement of the star's rotation period. Inspection of the 2010 photometric segment of
 Figure \ref{fig:phot1} clearly shows the overall 60-day modulation that generates the periodogram peak.
 Departures from perfect periodicity are presumably caused by the evolution of the spot activity on the surface of the star.

 Finally, we search for transits of GJ~687~b by first removing the spot
 variability from each of the yearly light curves. We do this by successively
 subtracting multiple frequencies from each yearly light curve using the
 method described in \citet{Henry01}. We removed three to six frequencies
 from each light curve until each set of residuals approached the precision
 of a single observation. The residuals from all five observing seasons are
 plotted in the top panel of Figure \ref{fig:phot3}, phased with the 38.14-day best-fit planetary orbital period and a time of mid transit computed from the orbital parameters. The vertical bar represents the 0.0022 mag standard deviation
 of the residuals from their mean, very close to the measurement precision
 given above. A sine fit to the phased data gives a formal semi-amplitude
 of just $0.00011~\pm~0.00012$~mag. Since none of the frequencies removed
 from the yearly light curves were similar to the orbital frequency or its
 harmonics, this result limits any periodic brightness variability of the
 star on the observed radial velocity period to a very small fraction of one
 milli-magnitude (mmag). This rules out the possibility that the 38.14-day
 radial velocity variations in GJ~687 are induced by stellar activity, as
 has been documented in somewhat more active stars, for instance, by
 \citet{Queloz01}, \citet{Paulson04}, and \citet{Boisse2012}. Instead, this
 lack of photometric variability confirms that the radial velocity variations
 in GJ~687 result from true planetary reflex motion.

 The photometric observations within $\pm0.13$P of mid-transit are replotted
 with an expanded scale in the bottom panel of Figure \ref{fig:phot3}. The solid curve
 shows the predicted phase, depth (assuming Neptune-like density), and duration of a central transit, computed
 from the stellar radius in Table \ref{tab:stellarparams} and the orbital elements in Table \ref{tab:fit}. The horizontal error bar under the predicted transit time gives the $\pm1\sigma$
 uncertainty in the timing of the transit. The photometric observations when filtered using the \citet{Henry01} procedure described above, and when folded at the $P=38.14$ day best-fit period for the planet, give no indication that transits occur. We note, however, that the Markov Chain models generate a five-day window for possible transits, and so a more conservative approach is also warranted. In Figure \ref{fig:transit}, we plot the unfiltered photometric data, indicating the range of photometric points that potentially could have been affected by transits were they to occur. Because of uncertainties in the orbit, the potential transit duration, the potential size of the planet, and the error in the photometric filtering, we recommend that continued photometric monitoring be carried out to confirm that transits do not occur.

 \section{Metallicity}

 Gliese 687 appears to have a slightly sub-solar metallicity. \citet{Rojas-Ayala12} use Na I, Ca I, and H$_{2}$O-K$2$ calibrations to estimate [Fe/H]=-0.09 for Gliese 687, whereas the M-dwarf metallicity calibration of Schlaufman \& Laughlin 2010 yields a value [Fe/H]=-0.02.

 The connection between the detectable presence of a giant extrasolar planet and the metallicity of the host star was noticed soon after the first extrasolar planets were detected \citep{Gonzalez97}, and has been studied in many previous works, see, e.g. \citet{FischerValenti05, Sousaetal11}. For M dwarfs, recent work, such as that by \citet{Neves13}, suggests that the giant planet stellar metallicity correlation holds robustly for M-dwarf primaries, but that for planets with mass, $M_{\rm p}\lesssim20\,M_{\oplus}$, no correlation is found with host star metallicity, and indeed, \citet{Neves13} report a hint of {\it anti}-correlation between the presence of a low-mass planet and host star $[{\rm Fe}/{\rm H}]$. Our detection of a Neptune-mass companion to Gliese 687, and our Lick-Carnegie database of Doppler measurements of M dwarf stars provides an opportunity to revisit this topic.

\begin{figure}
 \plotone{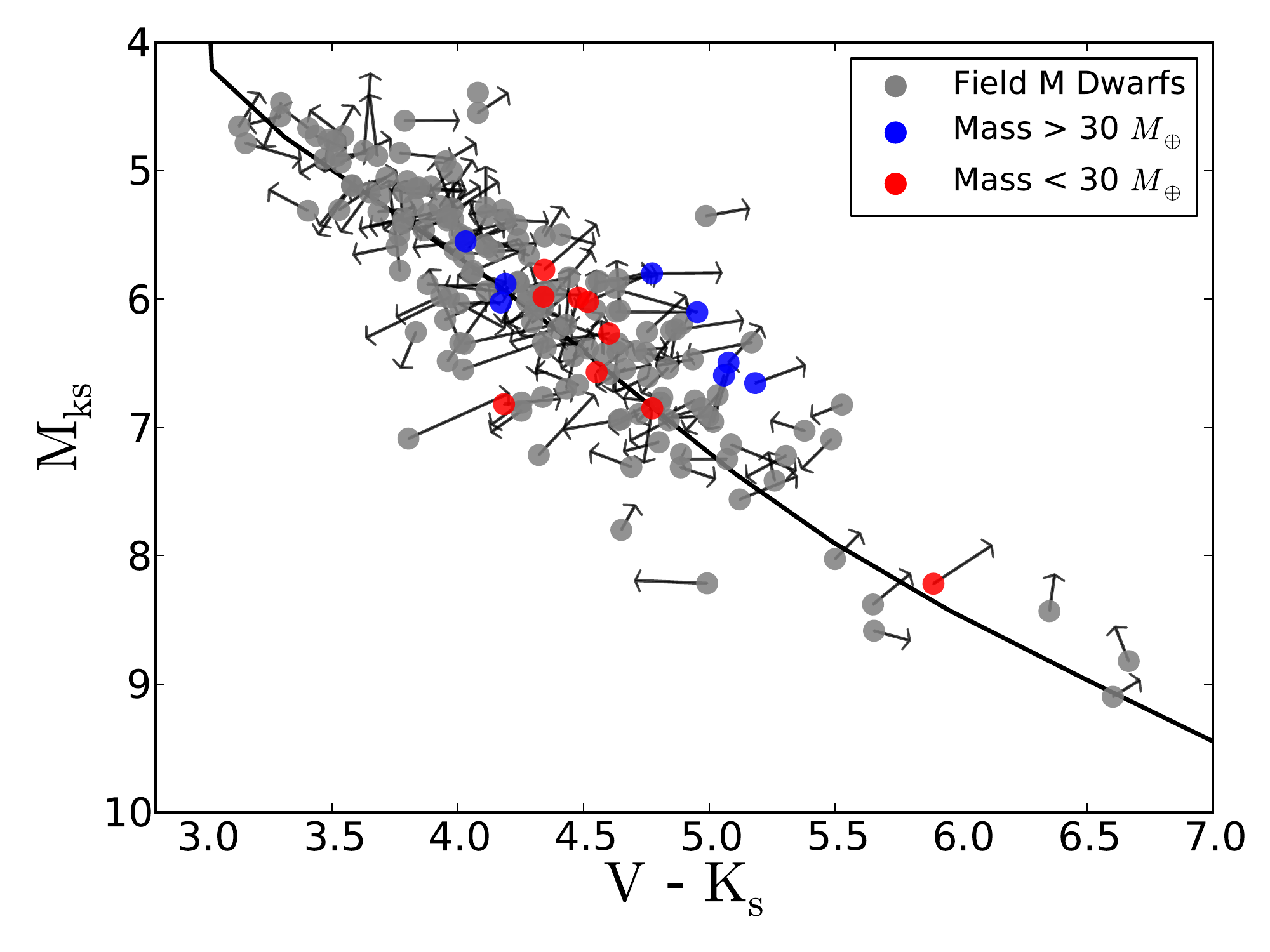}
 \plotone{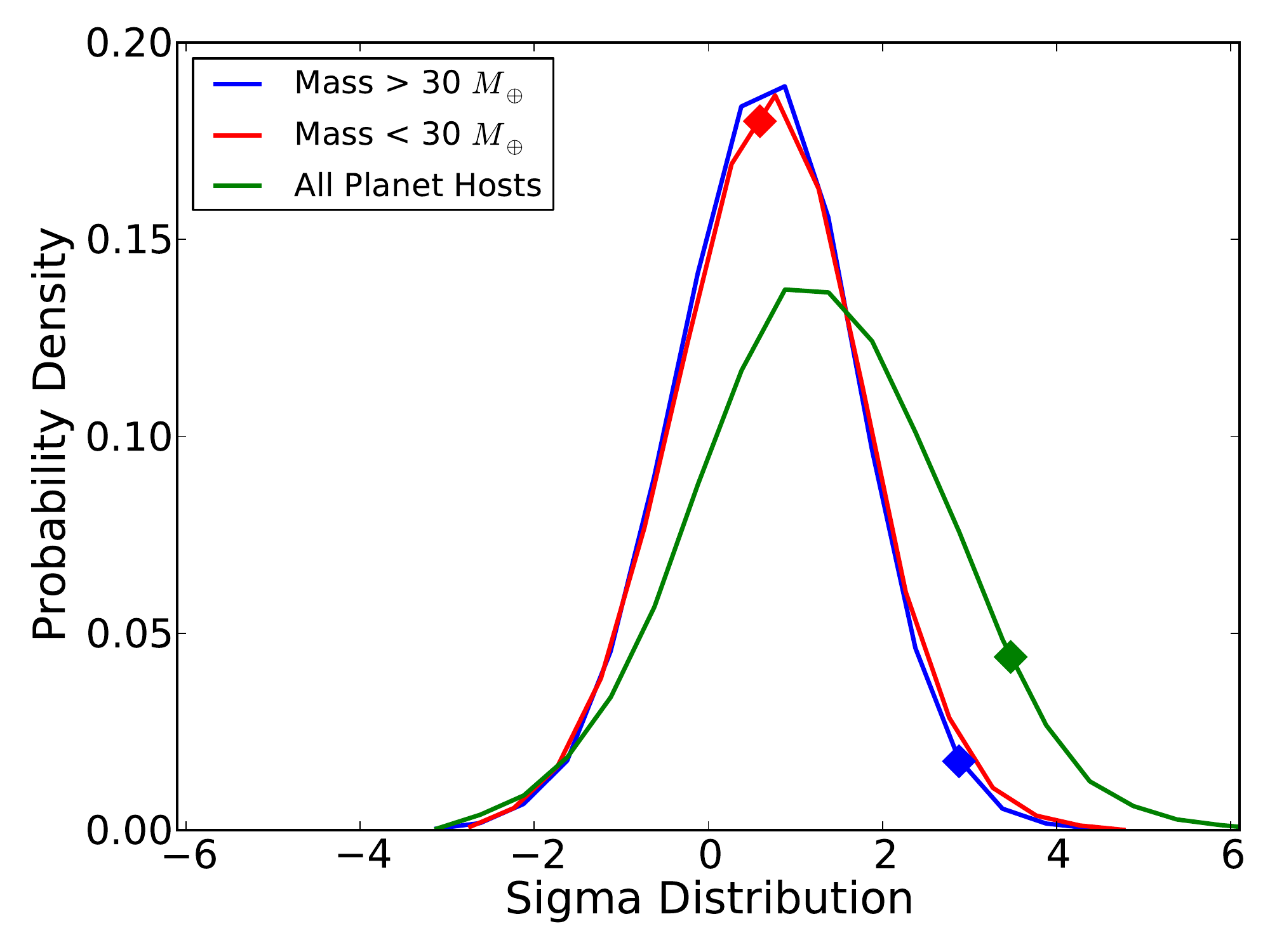}
 \caption{ {\it Top panel} Location of the 142 M dwarfs from the Lick-Carnegie radial velocity survey. Stars known to host Jupiter-mass planets are plotted in blue, those known to host twice-Neptune Msin(i) (or smaller) planets are plotted in red and non-planet hosting survey M dwarfs are plotted in grey. The field M dwarf main sequence from JA09 is shown as a black line and the arrows affixed to each point represent that survey star's proper motion. {\it Bottom panel} Distributions generated via Monte Carlo simulations of the cumulative sample distance of field M dwarfs from the M dwarf main sequence used by Johnson and Apps 2009. The points plotted on top of each curve in the bottom panel represent the actual cumulative distance from the MS for our planet hosting and field star samples.}
 \label{fig:metallicity}
 \end{figure}

 Our database of radial velocity observations taken at the Keck Telescope contains 142 M-type stars with the necessary spectral information to assess metallicity, 17 of which are known to host planets published in the peer-reviewed literature. We break the planet-hosting stars into two subgroups based on their masses - stars with $M\sin(i)$ planets less than 30 $M_{Earth}$ are described as Neptune hosting while stars with $M\sin(i)$ planets greater than 30 $M_{Earth}$ are listed as Jupiter hosting. We replicate the procedure of \citet{Schlaufman10} and examine how horizontal distance from a field M dwarf main sequence in a $M_{K_{s}}$ vs.~$(V - K_{s})$ color-magnitude diagram (CMD) correlates with metallicity, as noted e.g., by \citet{Baraffe98}. The top panel of Figure \ref{fig:metallicity} displays all of the Lick-Carnegie survey M dwarf stars plotted in $M_{K_{s}}$ vs.~$(V - K_{s})$, with grey dots denoting survey stars without known planets, red dots denoting survey stars that host ``Neptune-mass'' planets and blue dots representing the survey stars that host ``Jupiter-mass'' planets. It can be seen that most planet hosting stars fall to the right of the field M dwarf main sequence presented in \citet{Johnson09} (black line), which is taken to be a [Fe/H]=0.017 isometallicity contour in this CMD. In order to quantify the likelihood that a star's horizontal distance from the isometallicity contour is related to its propensity to host planets, we compare the distances for our actual planet-hosting stars with randomly drawn samples from the collection of M dwarfs in the survey.

 We characterize the position of each M dwarf by obtaining V-band and $K_{s}$ photometry and then using them to calculate the distance statistic $\Sigma$ :

 \begin{equation}
 \Sigma = \sum_{i=1}^{n} (V-K_{s})_{i} - (V - K_{s})_{iso}
 \end{equation}

 To determine if the $\Sigma$ of our known planet hosting subgroups is significant or, alternatively, if it could be produced by chance, we make use of a Monte Carlo simulation that calculates the cumulative sample distance of survey M dwarfs from the field M dwarf main sequence presented by Johnson and Apps 2009. For the simulation, we randomly select a subset of M dwarfs from the Lick-Carnegie field star list, setting the sample size equal to the number of M dwarfs known to host either Jovian or Neptune mass planets. Then we compute the cumulative horizontal distance of those stars from the field M dwarf MS, where stars to the right of the MS add their distance to the sum and stars to the left of the MS subtract their distance. We repeat this process 10,000 times to determine the distribution of cumulative horizontal distances from the MS given no correlation between whether the star hosts an exoplanet and its location in the $(V - K_{s}) - (M_{K_{s}})$ CMD.

 Our results show that for hosts of Jupiter mass planets, $\Sigma = 2.359$, which corresponds to a probability of $ p = 0.053^{+0.13}_{-0.04}$ that the stars' cumulative distance from the isometallicity contour occurred by chance. For the Neptune hosts, we find that $\Sigma = 1.113 $ leading to $ p = 0.452^{+0.24}_{-0.23} $, and for the combination of all planet hosts, we obtain $\Sigma = 3.473 $ or $ p = 0.0775^{+0.09}_{-0.05}$. The distributions resulting from the Monte Carlo simulation and the locations of the actual planet hosting stars $\Sigma$ values are displayed in the bottom panel of Figure \ref{fig:metallicity}. The points plotted on top of each curve in Figure \ref{fig:metallicity} represent the actual cumulative distance of our planet hosting star samples from the field M dwarf MS. Our results thus indicate that the planet-metallicity correlation is robust for M-dwarf hosts of planets with ${\rm M}\,>\,30{\rm M}_{{\oplus}}$, but that at smaller masses there is, at present, no evidence a correlation exists.

 \section{Planet Recovery}

 The Lick-Carnegie exoplanet survey and its predecessors have carried out a long-term monitoring program of the brightest M-dwarf stars in the sky. Our database of observations contains 159 stars that have more than 10 observations apiece, and which, additionally, have median internal uncertainty $\sigma<10\, {\rm m s^{-1}}$. Within this group, there is a subset with extensive data sets. For example, 11 stars have $N\, >\, 100$ observations and median internal uncertainties $\sigma<13\, {\rm m s^{-1}}$. A question of substantial interest, therefore, is the degree to which the observations taken to date have probed the true aggregate of planetary companions to the M-dwarf stars in our survey.

 The effort required to obtain the existing data has been substantial. Among the M-type stars alone, our database contains a total of 5,468 velocity measurements from Keck I, totaling 2,579,862 seconds (29.86 days) of on-sky integration. Overheads, including the acquisition of high S/N spectra, CCD readout time, and weather losses, add materially to this time investment. Furthermore, the distribution of total observing time allotted to the stars on the list has been highly uneven. Targets such as Gliese 436 and Gliese 876, which harbor planetary systems of particular interest, have received much more attention than the typical red dwarf in the survey. For example, Gliese 436 has 148 observations and Gliese 876 has 204 observations obtained with the Keck Telescope. The stars themselves also exhibit a range of chromospheric activity levels. The resulting star-to-star dispersion in ``stellar jitter" (tantamount to a measurement uncertainty, $\sigma_{jit}$) complicates the evaluation of threshold levels for $M\sin(i)$ as a function of orbital period to which planetary companions can be excluded.

 There are a variety of approaches to the measurement of false alarm probabilities (FAP) in the context of spectral analysis of unevenly sampled data. See, e.g. \citet{Baluev12} for a recent discussion. A very simple approach is described by \citet{Press}. For a gaussian random variable\footnote{Clearly, the generating function for typical radial velocity datasets has non-Gaussian (and unknown) error. False Alarm Probabilities must therefore be treated with great caution when evaluating the existence of a planet with $K \gtrsim \sigma_{\rm unc.}$. }, the probability distribution for obtaining a peak at frequency $\omega$ of Lomb-normalized power \citep{Scargle82}, $P_{\rm N}(\omega)$, is exponential with unit mean. If a data set drawn from measurements of a white noise (Gaussian) distribution supports measurement of $M$ independent frequencies, the probability that no peak exceeds power $z$ (the FAP) is ${\cal P}(P_{\rm N}>z)=1-(1-\exp^{-z})^{M}$.

\begin{figure}
 \plotone{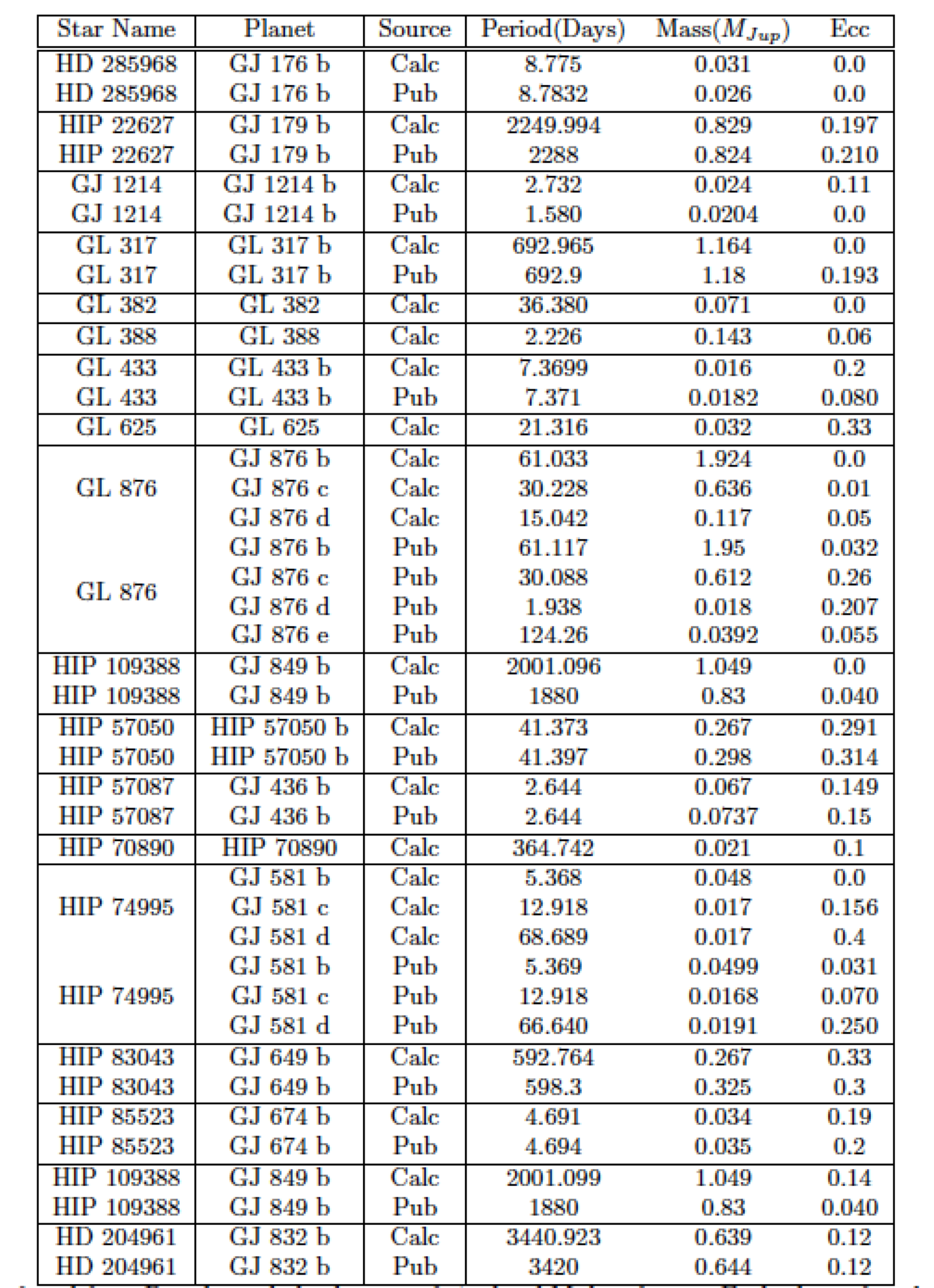}
 \caption{A table of recovered known extra solar planets orbiting M-Dwarf stars for which Doppler velocity measurements from the Keck telescope exist in the Lick-Carnegie database of observations. Published values (indicated with ``Pub'') are drawn from the compilation at www.exoplanets.org, accessed 2/14/2014. Also shown are the results obtained by our planet-finding algorithm (indicated with ``Calc''), when launched on a blind survey for planets.}
 \label{fig:MDwarfTable}
 \end{figure}

 We adopt a FAP of $10^{-4}$, calculated with the above method (and using Monte-Carlo simulations to determine $M$) as the generic threshold for attributing a given planetary signal to a given dataset. With this detectability threshold, we use the Systemic Console 2.0 software package \citep{Meschiari12} to determine the number of readily detectable planets in our M-dwarf data set. A ``readily detectable'' planet generates a signal that can be isolated algorithmically (and automatically) by straightforward periodogram analysis and Levenberg-Marquardt minimization. The results of this exercise are shown in Figure \ref{fig:MDwarfTable}, which locates signals corresponding to 19 previously published planets orbiting 14 separate M-dwarf primaries. Other than Gliese 667C, there are no stars on our 159-star list for which a planet has been published by another group, and for which the automated algorithm finds no planets. Regarding GJ~667C, 40 observations have been made at Keck, and these were used in a characterization of the GJ~667C system \citep{AngladaEscude12}, however the peak planetary signal for this set fell below our FAP threshold when utilizing only the Keck data. The bright planet-hosting red dwarfs Gliese 832, 3634, and 3470 all have declinations that are too far south to be observed from Mauna Kea, and HIP 79431 (RA 16h 12m 41.77s DEC -18$^{\circ}$ 52$^{\prime}$ 31.8$^{\prime \prime}$) is not on the list of M-dwarfs being monitored at Keck.

 \begin{figure*}
 \centering
 \includegraphics[width=7in]{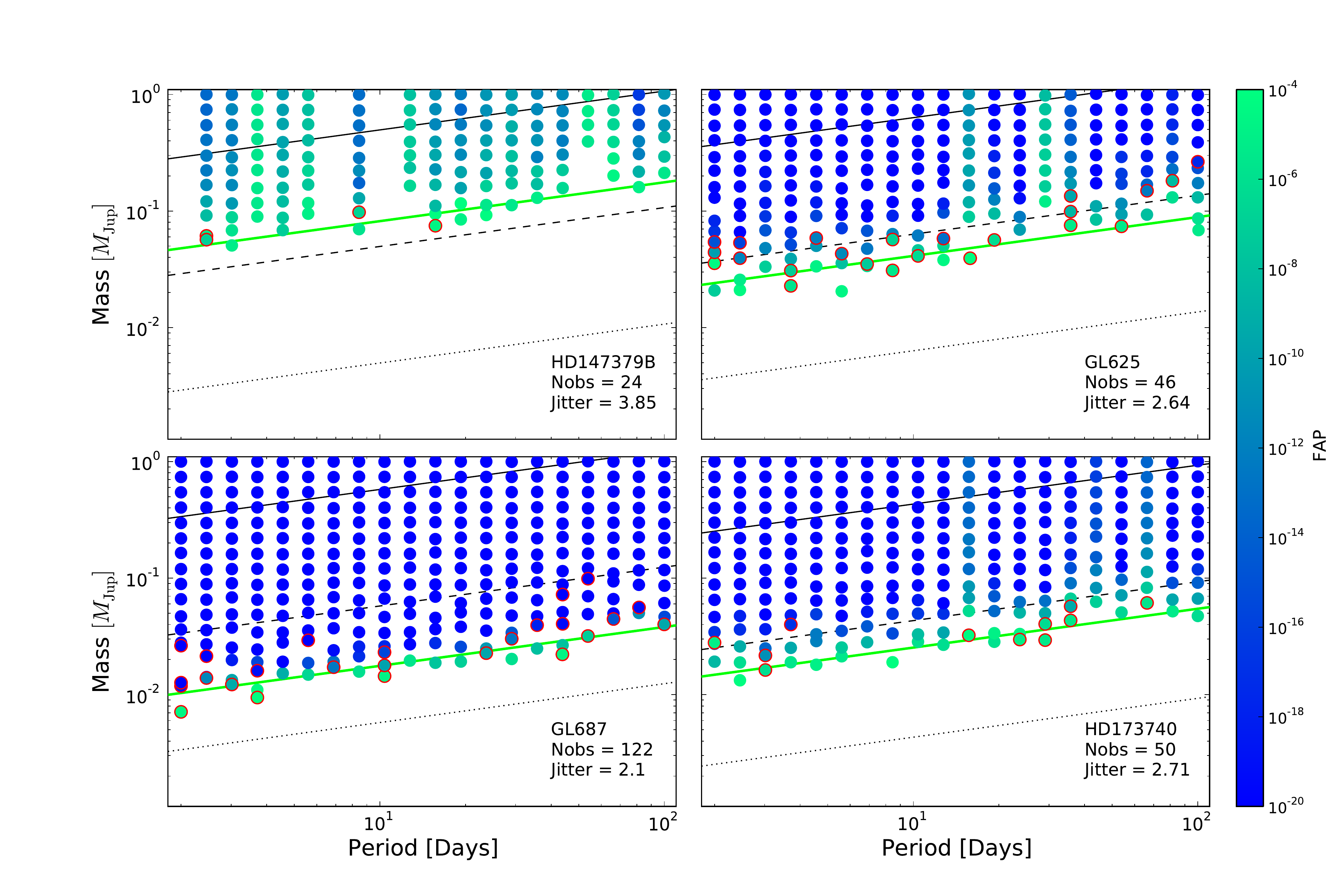}
 \caption{Example plots of our synthetic planet recovery around four M-dwarf stars. The points represent planets our algorithm found, colored by the false alarm probability for the initial detection. The black lines from bottom to top show radial velocity half amplitudes (K) of 1, 10, and 100 \ms. The green line is our minimum detectable K value.}
 \label{fig:massPerPanel}
 \end{figure*}

 \begin{figure}
 \plotone{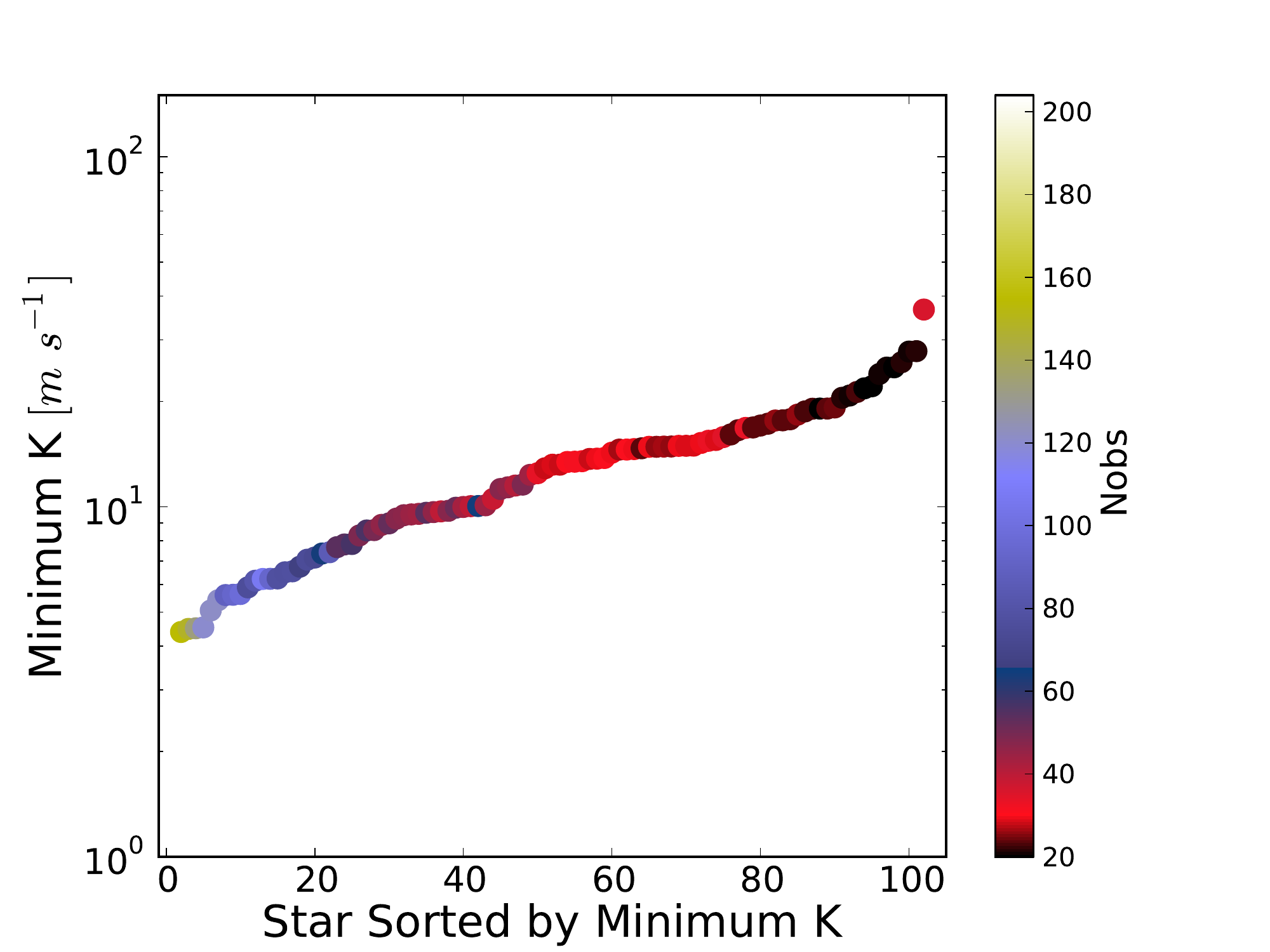}
 \plotone{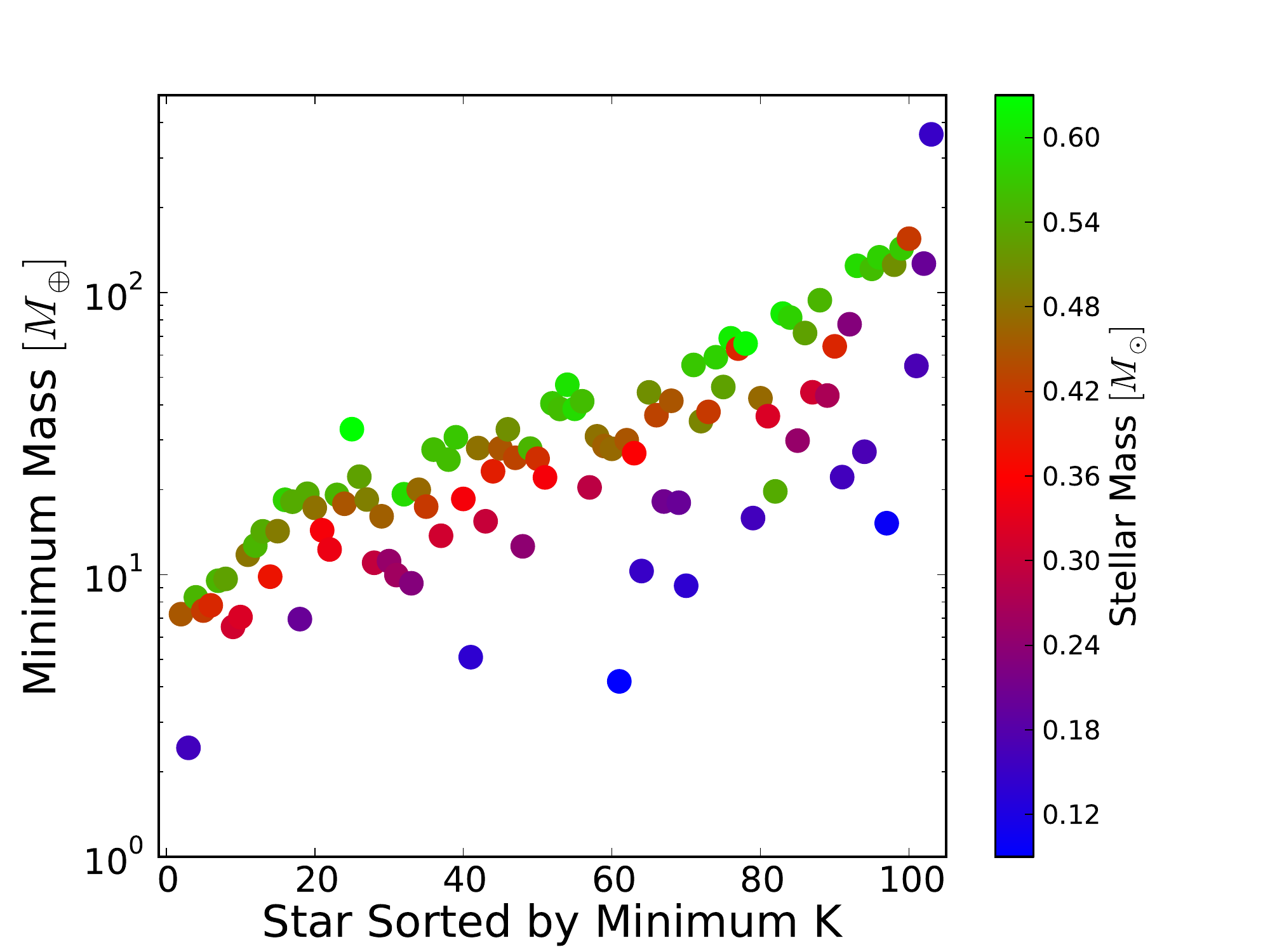}
 \caption{{\it Top panel} The minimum detectable $K$ value for each star in our M-dwarf collection. {\it Bottom panel} Assuming a minimum detectable $K$ for each M-dwarf star, if a planet was orbiting in that star's habitable zone, this is the minimum mass that planet could have and still be recovered by our method.}
 \label{fig:minK}
 \end{figure}

 The Kepler Mission's photometric data have been used to infer that small planets orbiting M-dwarfs are very common. For example, \citet{Dressing13} find an occurrence rate of 0.9 planets per star in the range $0.5\,R_{\oplus}<R_{\rm p}<4\,R_{\oplus}$ with ${\rm P} < 50$ days. Given the existence of this large number of small-radius planets, it is of interest to make a quantitative analysis of how deep into the expected population of super-Earth type planets suggested by the Kepler Mission the Keck Radial Velocity Survey has probed. To answer this question, we have created synthetic radial velocity data sets that contain test planets, and which conform with the timestamps, the internal measurement uncertainties, and the stellar properties (namely mass) for all 104 M-dwarfs under surveillance at Keck with at least 20 radial velocity observations. To address the error source arising from stellar jitter, $\sigma_{\rm jit}$, we use the median value provided in \citet{Wright05} of 3.9 ${\rm m\,s^{-1}}$ as the expected level of $\sigma_{\rm jit}$ for our M-dwarf stars. This value is then added in quadrature with the internal uncertainty and applied to the synthetic data set to create a more accurate representation of the system.

 For each of the 104 stars in the Keck survey, we have created 400 synthetic data sets. Each set contains a single planet. The planets are evenly spaced in log period from 2 to 100 days, and evenly spaced in log mass from 1 $M_{\oplus}$ to 1 $M_{\rm Jup}$. We assign a circular orbit to these test planets and assume $ i = 90^{\circ}$ and $\Omega=0^{\circ}$ in each case. We then calculate the radial velocity each of these planets would induce on a parent star. A Gaussian distribution with $\sigma^2 = {\sigma_{\rm internal}}^2 + {\sigma_{\rm jitter}}^2$ is used to perturb the predicted radial velocity value.

 Each of the $104\times400$ synthetic systems is passed to the planet search algorithm. Figure \ref{fig:massPerPanel} shows examples of the returned planets for 4 of the 104 stars in this experiment including the star of main interest here, GJ~687. In Figure \ref{fig:massPerPanel} the black lines from bottom to top represent constant $K={\left ( 2\pi G/P \right )}^{1/3} M_{\rm P} {M_{\star}}^{-2/3}$ values of 1, 10, and 100 \ms respectively. As expected, the detectablity thresholds lie roughly along lines of constant $K$. To determine the smallest $K$ value we could reliably detect for each star, we find the smallest value of $K$ for which a planet was found for at least 50\% of the chosen periods. This median value generates the green lines seen in the figure. The top panel of Figure \ref{fig:minK} shows these minimum $K$ values for each of the 104 stars that we tested. For clarity, the stars in this figure have been ordered by increasing minimum $K$, and are colored by the number of observations we have for each.

 If a test planet lies in its star's habitable zone, (defined as the semi-major axis at which the flux received by the planet is the solar constant received at Earth) we can ask how large the planet needs to be to be detectable by our radial velocity survey. Figure \ref{fig:minK} shows these threshold masses for each of the 104 M-dwarf stars which we analyzed. These stars maintain the ordering from the top panel, but now have been colored by the mass of the parent star. We see that while the Keck survey has probed substantially into the regime occupied by Neptune-mass planets, it has not made significant inroads into the super-Earth regime for periods that are of astrobiological interest.

 \section{Discussion}

 GJ~687 is the second planetary system to be detected using data from the APF telescope, with the first being HD 141399 b,c,d, and e \citep{Vogt14b}. APF has successfully navigated its commissioning stage, and, since Q2 2013, it has routinely acquired science-quality data that presents sub-m/s precision on known radial velocity standard stars \citep{Vogt14a}. In recent months, the degree of automation for APF has increased substantially. The facility currently works autonomously through an entire night's operations, calibration, and observing program. The APF and its accompanying high-resolution Levy spectrograph together form a dedicated, cost-effective, ground-based precision radial velocity facility that is capable of detecting terrestrial-mass planets at distances from their parent stars at which surface liquid water could potentially be present.

 Unlike other highly successful RV facilities, the APF uses neither image scrambling nor image slicing. With a peak efficiency of 15\% and typical spectral resolutions of $R\sim110,000$, the APF represents a critical new resource in the global quest to detect extrasolar planets. Initial speed comparisons indicate that in order to match the signal-to-noise acquired using the Keck telescope/HIRES Spectrograph combination, the APF needs only a factor of 6 increase in observing time. Since the amortized cost of a night on Keck is $\sim77\,\times$ more expensive than a night on the APF, and because 80\% of the APF's nights are reserved for exoplanetary work, the APF (with its sub-m/s precision and dedicated nightly cadence abilities) will likely provide key contributions to exoplanet detection and characterization in the coming years.

 Gliese 687~b's radial velocity half-amplitude, $K=6.4\pm0.5\,{\rm m s^{-1}}$, is substantially greater than the current state-of-the-art detection threshold for low mass planets. The lowest measured value for $K$ in the catalog of Doppler-detected extrasolar planets\footnote{www.exoplanets.org} stands at $K=0.51\,{\rm m s^{-1}}$ \citep{Dumusque12}. On the other hand, Gliese 687's status as one of the nearest stars to the Sun imbues it with a great deal of intrinsic interest. In our view, the relatively recent date for Gliese 687~b's detection can be attributed both to the substantial amount of stellar-generated radial velocity noise (as evidenced by Figures \ref{fig:sindex} and \ref{fig:phasedCurves}), but also to its location in Draco, high in the Northern Sky, where APF, along with HARPS North, are the only facilities that can routinely observe at sub-$1\,{\rm m s^{-1}}$ precision. (As evidenced by the data in this paper, Keck can observe at these high declinations, but at significantly higher expense in comparison to stars lying closer to the celestial equator.)

 Indeed, Gliese 687's stellar coordinates (R.A. 17h, 36m, DEC $+68^{\circ}$) place it very close to the north ecliptic pole, located at $RA=18$h, Dec$=+66^{\circ}$. This location flags it as a star of potentially great importance for the forthcoming NASA TESS Mission. As currently envisioned (and as currently funded), TESS is a two-year, all-sky photometric survey to be carried out by a spacecraft in a 27-day P/2 lunary resonant orbit. TESS will photometrically monitor $\sim500,000$ bright stars with a $<60\,{\rm ppm}$ 1-hour systematic error floor. (For reference, a central transit of the Sun by the Earth produces an 86 ppm transit depth.) The Northern Ecliptic Hemisphere will be mapped during the first year of the mission via a sequence of 13 sectors with 27 days of continuous observation per sector. These sectors overlap at the North Ecliptic Pole, and create an area of $\sim$1,000 square degrees (~1/50th of the sky) for which photometric baselines will approach 365 days. Gliese 687 lies at the center of this TESS ``overlap zone'' (which also coincides with JWST's continuous viewing zone). Because much longer time series are produced in the overlap zone, the highest-value transiting planets found by the mission will emerge from this part of the sky (along with the sister segment covering the South Ecliptic Pole).

 As mentioned above, however, the TESS overlap zone has received relatively little attention from the highest-precision Doppler surveys. About 10,000 target stars from the TESS Dwarf Star Catalog, all with V$<$12, are present in the overlap zone. This, of course, is far too many stars to survey with Doppler RV, but there appears to be substantial value inherent in monitoring the {\it brightest}, {\it nearest}, and {\it quietest} members of the cohort of TESS overlap stars. The latest estimates \citep{Mayor09, Mayor11, Batalha13, Petigura13} suggest that $\sim$50\% of main sequence stars in the solar vicinity harbor $M>M_{\oplus}$ planets with $P<100\,{\rm d}$. Assuming a uniform distribution in period between 5 and 100 days, the average transit probability for these planets is ${\cal P}\sim 2.5\%$, suggesting that of order ${\cal N}\sim0.5\times0.025\times10000~\sim~125$ low mass transiting planets (and systems of transiting planets) will be detected by TESS within the overlap zone. Of these, a small handful, of order 5 systems total (and perhaps, with probability ${\cal P}_{\rm transit}=1.2$\%, including Gliese 687~b) will garner by far the most attention from follow-up platforms such as JWST, due to their having optimally bright parent stars.

 Our detection of Gliese 687~b suggests that by starting now, with a systematic program of Doppler observations of a target list of $\sim$200 carefully vetted G, K \& M dwarf stars with V$\sim$7.5 to V$\sim$10.5 in the 1000-square degree TESS overlap zone, APF can ensure that a precise multi-year Doppler velocity time series will exist for the most important TESS planet host stars at the moment their transiting planets are discovered.

 \acknowledgments
 GL acknowledges support from the NASA Astrobiology Institute
 through a cooperative agreement between NASA Ames Research Center and the University of California at Santa Cruz. SSV gratefully acknowledges support from NSF grants AST-0307493 and AST-0908870.
 RPB gratefully acknowledges support from NASA OSS
 Grant NNX07AR40G, the NASA Keck PI program,
 and from the Carnegie Institution of Washington.
 S.M. acknowledges support from the W. J. McDonald Postdoctoral Fellowship.
 GWH acknowledges support from NASA, NSF, Tennessee State University, and
 the State of Tennessee through its Centers of Excellence program.
 The work herein is based on
 observations obtained at the W. M. Keck Observatory, which is operated jointly by the University of California and
 the California Institute of Technology, and we thank the UC-Keck, UH, and NASA-Keck Time Assignment Committees for their
 support. This research has made use of the Keck Observatory Archive (KOA), which is operated
 by the W. M. Keck Observatory and the NASA Exoplanet Science Institute (NExScI), under contract with
 the National Aeronautics and Space Administration. We thank those who collected the data in the KOA
 including: D. Fischer, G. Marcy, J. Winn, W. Boruki, G. Bakos, A. Howard, and T. Bida. We also wish to extend our
 special thanks to those of Hawaiian ancestry on whose sacred mountain
 of Mauna Kea we are privileged to be guests. Without their generous hospitality, the Keck observations
 presented herein would not have been possible. This research has made use of the SIMBAD database,
 operated at CDS, Strasbourg, France. This paper was produced using $^BA^M$.

 \vspace{\baselineskip}
 {\it Facilities:} \facility{Keck (HIRES)}, \facility{Automated Planet Finder (Levy Spectrometer)}, \facility{Hobby-Eberly (High-Resolution-Spectrograph)} .

 \clearpage

 \interfootnotelinepenalty=10000

 \appendix
 \section[Doppler Radial Velocity Observations]{Doppler Radial Velocity Observations\footnote{O\lowercase{bservations are corrected to the solar system's barycentric reference frame.}}}

 % Tables of RV data

 \LongTables
 \begin{deluxetable}{ccc}
 \tablecaption{HIRES/Keck radial velocities for GJ~687
 \label{tab:rvdata_HIRES/Keck}}
 \tablecolumns{3}
 \tablehead{{JD}&{RV [m/s]}&{Uncertainty [m/s]}}
 \startdata
 2450603.965 & -15.930 & 1.770  \\
2450956.065 & -4.740 & 2.040  \\
2450982.977 & -12.880 & 1.560  \\
2451013.879 & -15.550 & 1.610  \\
2451312.036 & -6.870 & 1.910  \\
2451368.798 & -17.330 & 1.790  \\
2451439.745 & -25.500 & 2.170  \\
2451704.957 & -14.920 & 1.810  \\
2452007.029 & -3.280 & 1.970  \\
2452009.077 & -7.440 & 2.020  \\
2452061.913 & 7.450 & 2.040  \\
2452062.935 & 4.490 & 1.940  \\
2452094.853 & 2.450 & 1.550  \\
2452096.894 & 4.560 & 1.850  \\
2452097.982 & 5.180 & 1.610  \\
2452127.919 & 1.940 & 2.200  \\
2452133.737 & 2.830 & 2.010  \\
2452160.872 & -6.120 & 2.170  \\
2452161.821 & -6.840 & 2.370  \\
2452162.787 & -8.300 & 2.370  \\
2452445.988 & 17.880 & 2.100  \\
2452537.743 & -4.900 & 2.340  \\
2452713.113 & 8.860 & 1.590  \\
2452806.027 & -4.580 & 1.990  \\
2452850.901 & 6.090 & 2.040  \\
2453179.985 & 12.360 & 1.730  \\
2453479.066 & 8.000 & 1.210  \\
2453549.857 & -2.830 & 0.880  \\
2453604.881 & -2.530 & 1.220  \\
2453838.109 & 3.550 & 1.220  \\
2453932.905 & 3.890 & 1.380  \\
2453960.873 & 3.880 & 1.630  \\
2453961.820 & -0.510 & 1.800  \\
2453981.789 & -14.390 & 1.190  \\
2453982.904 & -1.880 & 1.900  \\
2453983.831 & -8.010 & 1.220  \\
2453984.887 & -0.360 & 1.290  \\
2454248.023 & 7.800 & 2.090  \\
2454248.990 & 8.410 & 2.030  \\
2454249.945 & -3.960 & 1.730  \\
2454252.032 & -12.260 & 1.870  \\
2454255.924 & -12.790 & 1.280  \\
2454277.851 & 8.200 & 1.500  \\
2454278.896 & 11.300 & 1.700  \\
2454279.933 & 11.850 & 1.580  \\
2454294.903 & -10.850 & 1.260  \\
2454304.888 & -4.110 & 1.570  \\
2454306.037 & 2.450 & 1.600  \\
2454307.015 & -2.050 & 1.500  \\
2454308.072 & 5.540 & 1.600  \\
2454309.050 & 7.370 & 1.810  \\
2454310.042 & 8.730 & 1.720  \\
2454311.024 & 1.600 & 1.480  \\
2454312.017 & 0.860 & 1.590  \\
2454312.875 & 2.350 & 1.450  \\
2454313.875 & -1.170 & 1.320  \\
2454314.908 & 6.220 & 1.630  \\
2454318.921 & 5.360 & 1.560  \\
2454335.826 & -13.560 & 1.940  \\
2454338.888 & -16.490 & 1.790  \\
2454339.883 & -9.590 & 1.780  \\
2454343.791 & -8.170 & 1.950  \\
2454396.717 & 3.080 & 2.390  \\
2454397.729 & -4.380 & 2.130  \\
2454548.052 & 12.600 & 1.330  \\
2454549.092 & 3.010 & 1.280  \\
2454633.963 & -6.780 & 1.570  \\
2454634.903 & -5.330 & 1.650  \\
2454635.946 & -2.500 & 1.790  \\
2454636.901 & 2.170 & 1.650  \\
2454637.939 & -5.830 & 1.610  \\
2454638.838 & -5.010 & 1.450  \\
2454639.998 & 3.970 & 1.850  \\
2454641.954 & 2.760 & 1.540  \\
2454674.860 & -0.820 & 1.440  \\
2454688.892 & -4.450 & 1.710  \\
2454689.917 & 6.340 & 1.890  \\
2454717.853 & 4.220 & 2.000  \\
2454718.914 & -2.870 & 2.070  \\
2454719.866 & 1.210 & 2.040  \\
2454720.871 & 0.570 & 2.140  \\
2454721.879 & 4.620 & 2.040  \\
2454722.797 & 5.810 & 2.210  \\
2454723.814 & 2.610 & 2.120  \\
2454724.848 & 1.350 & 1.990  \\
2454968.017 & 5.210 & 0.970  \\
2455016.035 & 0.730 & 1.370  \\
2455022.084 & 1.680 & 1.670  \\
2455023.044 & 6.360 & 1.070  \\
2455024.823 & -11.730 & 0.920  \\
2455049.825 & 0.220 & 1.090  \\
2455050.858 & 3.990 & 1.650  \\
2455051.772 & 6.360 & 1.790  \\
2455052.780 & 8.540 & 1.930  \\
2455053.925 & -12.540 & 1.510  \\
2455143.692 & -3.080 & 2.620  \\
2455167.704 & 3.210 & 1.710  \\
2455259.119 & 3.760 & 1.610  \\
2455260.144 & 0.000 & 1.470  \\
2455371.991 & -5.030 & 1.070  \\
2455407.955 & -1.480 & 1.060  \\
2455463.718 & -8.590 & 1.520  \\
2455516.696 & -1.120 & 1.740  \\
2455518.722 & -3.780 & 2.320  \\
2455609.135 & 11.250 & 1.310  \\
2455638.121 & -3.930 & 0.910  \\
2455639.143 & -2.250 & 0.910  \\
2455665.046 & -5.130 & 1.190  \\
2455670.082 & -8.090 & 1.290  \\
2455721.057 & 6.840 & 1.850  \\
2455825.832 & 3.030 & 2.280  \\
2455840.748 & 3.820 & 1.390  \\
2456027.098 & 11.390 & 1.940  \\
2456116.938 & 16.010 & 1.130  \\
2456117.906 & 14.270 & 1.050  \\
2456329.167 & 5.420 & 1.970  \\
2456432.892 & -0.700 & 1.300  \\
2456433.935 & 1.610 & 1.340  \\
2456548.855 & -2.340 & 1.490  \\
2456549.793 & -7.780 & 1.430  \\
2456550.849 & -3.840 & 2.150  \\
2456551.724 & -2.520 & 1.570  \\

 \enddata
 \end{deluxetable}
 \LongTables
 \begin{deluxetable}{ccc}
 \tablecaption{Hobby-Eberly Telescope radial velocities for GJ~687
 \label{tab:rvdata_Hobby-Eberly Telescope}}
 \tablecolumns{3}
 \tablehead{{JD}&{RV [m/s]}&{Uncertainty [m/s]}}
 \startdata
 2452394.962 & -2.010 & 2.220  \\
2452395.925 & -3.600 & 2.600  \\
2452396.936 & -1.400 & 2.870  \\
2452400.920 & 3.800 & 2.880  \\
2452403.869 & 1.370 & 2.290  \\

 \enddata
 \end{deluxetable}
 \LongTables
 \begin{deluxetable}{ccc}
 \tablecaption{APF radial velocities for GJ~687
 \label{tab:rvdata_APF}}
 \tablecolumns{3}
 \tablehead{{JD}&{RV [m/s]}&{Uncertainty [m/s]}}
 \startdata
 2456484.738 & 11.650 & 0.910  \\
2456486.839 & 9.500 & 0.950  \\
2456488.845 & 11.640 & 1.120  \\
2456490.726 & 8.300 & 1.260  \\
2456492.816 & 7.020 & 0.900  \\
2456494.776 & 5.260 & 0.750  \\
2456508.764 & -8.650 & 0.780  \\
2456510.800 & -9.020 & 0.800  \\
2456513.772 & -8.920 & 0.860  \\
2456521.823 & -2.120 & 0.730  \\
2456522.790 & -2.920 & 0.930  \\
2456533.797 & 2.910 & 0.820  \\
2456551.743 & -8.410 & 0.810  \\
2456552.717 & -6.660 & 0.970  \\
2456553.727 & -7.610 & 0.850  \\
2456554.720 & -4.640 & 0.890  \\
2456564.762 & 1.530 & 0.940  \\
2456568.692 & 0.460 & 1.140  \\
2456571.657 & 0.580 & 0.940  \\
2456622.032 & -9.750 & 1.280  \\
2456677.921 & 2.390 & 0.980  \\
2456681.011 & 4.660 & 1.360  \\

 \enddata
 \end{deluxetable}

 % Bibliography
 \clearpage
% \bibliographystyle{apj}
% \bibliography{biblio}

 \end{document}